\documentclass{IEEEtran}
\usepackage{amssymb}
\usepackage{amsfonts}
\usepackage{amsmath}
\usepackage{algpseudocode}
\usepackage{algorithm}
\usepackage{caption}
\usepackage{bm}
\usepackage{graphicx,subfigure,cite,multicol,multirow,diagbox,booktabs,array}
\usepackage{stfloats}

\usepackage{color}
\usepackage{float}
\usepackage{subfig}
\usepackage{makecell}
\usepackage{titlesec}
\makeatletter

\newcommand{\Rmnum}[1]{\expandafter\@slowromancap\romannumeral #1@}
\makeatother
\ifCLASSINFOpdf
\else
\fi
\begin{document}	

\title{
Machine Learning-based Near-field Emitter Location Sensing via Grouped Hybrid Analog and Digital XL-MIMO Receive Array
}
\author{Jiatong Bai, Yifan Li, Feng Shu,~\emph{Member},~\emph{IEEE}, Kang Wei, Cunhua Pan,~\emph{Senior Member},~\emph{IEEE}, Yongpeng Wu,~\emph{Senior Member},~\emph{IEEE}, Yaoliang Song, and Jiangzhou Wang,~\emph{Fellow},~\emph{IEEE}

\thanks{Corresponding author: Feng Shu, Yifan Li}
\thanks{Y. Li and Y. Song are with the School of Electronic and Optical Engineering, Nanjing University of Science and Technology, Nanjing 210094, China. (e-mail: liyifan97@foxmail.com).}
\thanks{F. Shu and J. Bai are with the School of Information and Communication Engineering, Hainan University, Haikou 570228, China. (e-mail: shufeng0101@163.com).}
\thanks{Kang Wei is with the Department of Computing, The Hong Kong Polytechnic University, Hong Kong 100872, China (e-mail: kangwei@polyu.edu.hk).}
\thanks{C. Pan is with the National Mobile Communications Research Laboratory, Nanjing 210096, China. (e-mail:
	cpan@seu.edu.cn).}
\thanks{Yongpeng Wu is with the Department of Electronic Engineering, Shanghai Jiao Tong University, Minhang 200240, China (e-mail: yongpeng.wu@sjtu.edu.cn).}
\thanks{J. Wang is with the School of Engineering, University of Kent, Canterbury CT2 7NT, U.K. (e-mail: j.z.wang@kent.ac.uk).}
\vspace{-1em}
}
\maketitle

\begin{abstract}
As a green MIMO structure, the partially-connected hybrid analog and digital (PC-HAD) structure has been widely used in the far-field (FF) scenario for it can significantly reduce the hardware cost and complexity of large-scale or extremely large-scale MIMO (XL-MIMO) array. 
Recently, near-field (NF) emitter localization including direction-of-arrival (DOA) and range estimations has drawn a lot of attention, but is rarely explored via PC-HAD structure. In this paper, we first analyze the impact of PC-HAD structure on the NF emitter localization and observe that the phase ambiguity (PA) problem caused by PC-HAD structure can be removed inherently with low-latency in the NF scenario. To obtain the exact NF DOA estimation results, we propose a grouped PC-HAD structure, which is capable of dividing the NF DOA estimation problem into multiple FF DOA estimation problems via partitioning the large-scale PC-HAD array into small-scale groups. An angle calibration method is developed to address the inconsistency among these FF DOA estimation problems. Then, to eliminate PA and improve the NF emitter localization performance, we develop three machine learning (ML)-based methods, i.e., two low-complexity data-driven clustering-based methods and one model-driven regression method, namely RegNet. Furthermore, the Cramer-Rao lower bound (CRLB) of NF emitter localization for the proposed grouped PC-HAD structure is derived and reveals that localization performance will decrease with the increasing of the number of groups. The simulation results show that the proposed methods can achieve CRLB at different SNR regions, the RegNet has great performance advantages at low SNR regions and the clustering-based methods have much lower computation complexity.
\end{abstract}

\section{Introduction}
With the development of sixth generation (6G) mobile communications, the requirements of larger array apertures and higher frequency bands drive the application of technologies such as extremely large-scale MIMO (XL-MIMO), Terahertz (THz), reconfigurable intelligent surface (RIS), etc. \cite{zhao20246g}.  However, the larger array aperture and shorter wavelength also induce some changes in the modeling of propagation patterns. Different from the conventional MIMO, when the number of antennas in an XL-MIMO system increases to hundreds, the emitters or mobile stations are more likely located inside the near-field (NF) region rather than far-field (FF) \cite{zhou2015spherical}, and the spherical wave propagation is applicable \cite{yin2017scatterer}. Therefore, this leads us to focus on the research of NF problems in the XL-MIMO systems.

NF localization is a fundamental technology in NF communication, and plays an important role in applications like channel estimation\cite{cui2022channel, 10664591, zhao2025}, integrated sensing and communication (ISAC) \cite{mao2022waveform}, physical layer security \cite{zhang2024physical}, etc.. The existing NF localization methods can be divided into two categories, the first is based on the conventional estimation methods. In \cite{huang1991near}, the conventional multiple signal classification (MUSIC) and maximum likelihood estimator (MLE) were first applied for the NF passive localization. The work in \cite{liang2009passive} proposed a two-stage MUSIC algorithm for localization of mixed NF and FF sources. A modified MUSIC algorithm for NF localization was proposed in \cite{zhang2018localization}, which has a reduced dimension compared to 2D-MUSIC algorithm and achieves much lower complexity without losing performance. The second category of algorithms are on the basis of high-order statistics. The work in \cite{challa1995high} firstly proposed an NF localization algorithm by using fourth-order statistics. To further improve the performance, a second-order statistics-based weighted linear prediction algorithm was proposed in \cite{grosicki2005weighted}.

Recently, machine learning (ML) or deep learning (DL) - based methods have become popular in emitter localization because they can provide high estimation accuracy in both FF and NF scenarios. A deep neural network with a regression layer was adopted in \cite{liu2019deep} to address the problem of NF source localization. In \cite{cao2020complex}, a complex ResNet was designed for NF DOA estimation. The work in \cite{jang2024neural} introduced a neural network for joint optimization of beamformer and localization function to achieve NF channel estimation. For the mixed NF and FF scenarios, the work in \cite{su2021mixed} developed convolution neural networks (CNN) for sources classification and localization by using the geometry of symmetric nested array. Classical clustering methods like $k$-means, density-based spatial clustering of applications with noise (DBSCAN), and global maximum cos\_similarity clustering are also applied to FF DOA estimation\cite{shu2024newheterogeneoushybridmassive, 10892212, bai2024passive}. Based on the existing works, ML-based methods can provide significant performance gains in the resolve of localization problems.

Another important factor that can affect the performance of NF localization is the structure of receive array.
Conventional large-scale MIMO receive array with fully-digital (FD) structure can provide pretty high resolution gain for the emitter localization, but the circuit cost and energy consumption of corresponding radio-frequency (RF) chains and analog to digital conversion will also significantly increase. Therefore, in consideration of green communication and cost, hybrid analog and digital (HAD) structure is a better choice for achieving a balance between cost and performance \cite{zhang2005variable}. HAD structures are mainly divided into three categories: fully-connected (FC), partially-connected (PC) and switches-based (SW) \cite{ioushua2019family,ratnam2018hybrid,sohrabi2016hybrid}. 
 The HAD structures have been widely considered in FF DOA estimation\cite{zhang2021direction,chuang2015high,shu2018low}, but their ability wes rarely explored in NF situation. As two key techniques in XL-MIMO communication and 6G,  it is essential to investigate the challenges that arise from integrating HAD structures with NF scenarios, so we first consider an NF localization problem via PC-HAD structure in this work. 

As a green structure, PC-HAD can save many resources compared to FD structure, but it also introduce a new problem of phase ambiguity (PA) in DOA estimation \cite{shu2018low,zhang2021direction}. In \cite{shu2018low}, a two-stage method was designed to remove the PA, but this method has low efficiency. Then in order to improve the time efficiency of PA elimination algorithm, a rapid method was proposed in \cite{zhan2024rapid} which is in sacrifice of a few performance. Finally, a new heterogeneous PC-HAD structure was proposed in \cite{shu2024newheterogeneoushybridmassive} to achieve high-performance PA elimination with a low-latency. However, the PA problems considered in above works are all in the FF scenario. When the scenario turn to NF, these methods will not be available,  so it is necessary for us to reconsider how to resolve the PA problem in the NF situation.

Up to now, the problem of NF localization is rarely explored via PC-HAD structure, and the corresponding PA problem also have not been resolved. In this paper, we will focus on the investigation of these problems.
The main contributions of this work are summarized as follows:
\begin{enumerate}
	\item By analyzing the impact of PC-HAD structure on NF emitter localization, we find the PC-HAD structure has an endogenous ability of removing the PA with a low-latency due to the non-linear effect of NF model. Then in order to solve the NF DOA estimation problem, a grouped PC-HAD structure is developed. By dividing the large-scale receive array into multiple small-scale groups, the NF problem of estimating
	DOA within each group is viewed as a FF one, so the existing low-complexity and high-precision FF DOA estimation methods can be utilized. Furthermore, to address the inconsistency between the DOA estimation results of different groups, we also develop an angle calibration method. 
	\item To deal with the problem of PA and obtain the NF emitter position, two data-driven low-complexity clustering-based methods, i.e., minimum sample distance clustering (MSDC) and range scatter diagram-angle scatter diagram-based DBSCAN (RSD-ASD-DBSCAN), are developed. Firstly, MSDC is proposed based on the range-angle combination  distribution feature of samples in the generated candidate position set. Then by observing the different distribution features of samples in range and angle dimensions respectively, the RSD-ASD-DBSCAN method is developed to achieve higher localization precision.
	\item To further improve the performance of PA elimination and emitter localization, a model-driven regression network (RegNet) combining a multi-layer neural network (MLNN) and a single-layer perceptron is developed. The MLNN part simulates the nonlinear mapping process of PA elimination, and the perceptron fuses the true DOA solutions of all the groups to generate the final DOA estimation result. Then by combining the outputs of MLNN and perceptron, the range can also be estimated based on the principle of angle calibration. 
	\item To evaluate the performance of the proposed structure and methods, the closed-form expression of the corresponding CRLB for the grouped HAD structure is derived. The expression of CRLB reveals its value decreases with respect to the increasing of the number of subarrays in each group, which is in line with the simulation results. Then let CRLB be a benchmark for the proposed methods, we observe the proposed methods can all achieve CRLB. 
\end{enumerate}

The remainder of this paper is organized as follows. Section $\rm{\Rmnum{2}}$ introduces the system model. Section $\rm{\Rmnum{3}}$ describes the grouped PC-HAD structure and gives the corresponding DOA estimation method. In Section $\rm{\Rmnum{4}}$, two low-complexity clustering-based localization methods are proposed. The high-resolution RegNet is proposed in Section $\rm{\Rmnum{5}}$. Section $\rm{\Rmnum{6}}$ makes some discussions and analysis. Finally, simulation results and conclusions are given in Sections $\rm{\Rmnum{7}}$ and $\rm{\Rmnum{8}}$ respectively.

\textit{Notations:} Matrices, vectors and scalars are represented by letters of bold upper case, bold lower
case, and lower case, respectively. Signs $(\cdot)^T$, $(\cdot)^*$ and $(\cdot)^H$ denote transpose, conjugate and conjugate transpose. ${\rm{E}}[\cdot]$ denotes statistical expectation. ${\rm{tr}\{\cdot\}}$ is the trace of a matrix and ${\rm{diag}}\{\cdot\}$ represents a diagonal matrix. $\lVert\cdot\rVert$ is the Euclidean norm. $\Re\{\cdot\}$ and $\Im\{\cdot\}$ stand for the real part and imaginary part of a complex number. ${n\choose m}$ denotes the number of combinations of $m$ elements taken from $n$ different elements. $\boldsymbol{1}$ denotes an vector filled with element 1.

\vspace{-1em}
	\section{System Model}
\begin{figure}[t]
		\centering
		\includegraphics[width=0.3\textwidth]{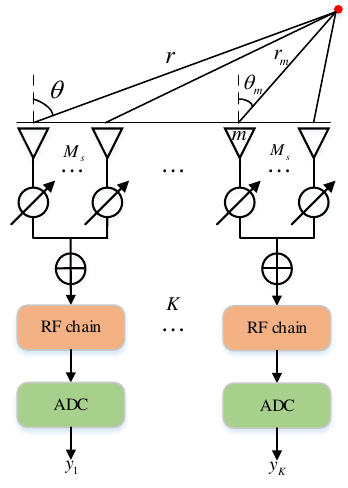}\\
		\caption{NF emitter localization model via PC-HAD receiver.}\label{system model}
		\vspace{-1em}	
	\end{figure}

	As shown in Fig. \ref{system model}, consider a NF narrowband signal impinges on a $M$-antennas ULA, the received signal model at $m$th antenna, $ m \in \{1, 2, \ldots, M\}$ is given by
	\begin{equation}
		x_m(t)=s(t)e^{j\varphi_{m}}+v_m(t),\label{y_m}
	\end{equation}
	where $s(t)$ denotes the signal waveform, $v_m(t)\sim \mathcal{CN}(0,\sigma_v^2)$ is the additive white Gaussian noise (AWGN), and $\varphi_{m}$ represents the phase difference caused by the signal propagation delay between the $m$th antenna and the reference point. Based on the uniform spherical model, $\phi_{m}$ is given by
	\begin{equation}
		\phi_{m}=\frac{2\pi}{\lambda}\left(\sqrt{r^2+(m-1)^2d^2-2(m-1)dr\sin\theta}-r\right),\label{phase difference}
	\end{equation}
	where $\theta\in[-\pi/2,\pi/2]$ and $r\in[0.62(D^3/\lambda)^{1/2},2D^2/\lambda]$ are angle and range parameters to be estimated, $\lambda$ denotes signal wavelength, $d=\lambda/2$ is the interval between two adjacent antennas and $D$ denotes the array aperture. The range of $r$ is the so-called Fresnel region of the receive array.
	
	As the receive array is with PC-HAD architecture, it is partitioned into $K$ subarrays and each subarray is connected to an RF chain. So the output signal of the $k$th RF chain, $ k \in \{1, 2, \ldots, K\}$ is given as
	\begin{equation}
		\begin{aligned}
		y_k(t)=\mathbf{w}_k^H\mathbf{x}_k(t)=\frac{1}{\sqrt{M_s}}\sum_{l=1}^{M_s}s(t)e^{j(\varphi_{k,l}+\alpha_{k,l})}+v_{k,l}(t)e^{j\alpha_{k,l}},\label{y_k}
		\end{aligned}
	\end{equation}
where $M_s$ denotes the number of antennas in each subarray and $M=KM_s$, $\mathbf{w}_k=\frac{1}{\sqrt{M_s}}[e^{j\alpha_{k,1}},e^{j\alpha_{k,2}},\cdots,e^{j\alpha_{k,M_s}}]^T$ is the analog beamforming vector.	
Via combining the output signals of all the $K$ RF chains, (\ref{y_k}) is transformed to the following vector form :
	\begin{equation}
		\begin{aligned}
			\mathbf{y}(t)&=\left[y_1(t), y_2(t), \cdots,y_K(t) \right]^T\\
			&=\mathbf{W}^H\mathbf{a}(\theta,r)s(t)+\mathbf{W}^H\mathbf{v}(t),
		\end{aligned}
	\end{equation} 	
	where $\mathbf{W}$ is a block diagonal matrix defined as 
	\begin{equation}
		\mathbf{W}={\rm{diag}}\left\{\mathbf{w}_1,\mathbf{w}_2,\cdots,\mathbf{w}_K\right\}\in \mathbb{C}^{M\times K},\label{W}
	\end{equation} 
	$\mathbf{v}(t)=[v_0(t),v_1(t),\cdots,v_{M-1}(t)]^T$ represents the noise vector of the receive array and
	$\mathbf{a}(\theta,r)=\left[1,e^{j\varphi_2},\cdots,e^{j\varphi_M}\right]^T\in \mathbb{C}^{M\times 1}$
	is the NF array steering vector. As can be observed, different from the FF cases, the localization of NF emitters involves not only angle estimation but also range estimation. 
	
\begin{figure}[t]
	\centering
	\includegraphics[width=0.4\textwidth]{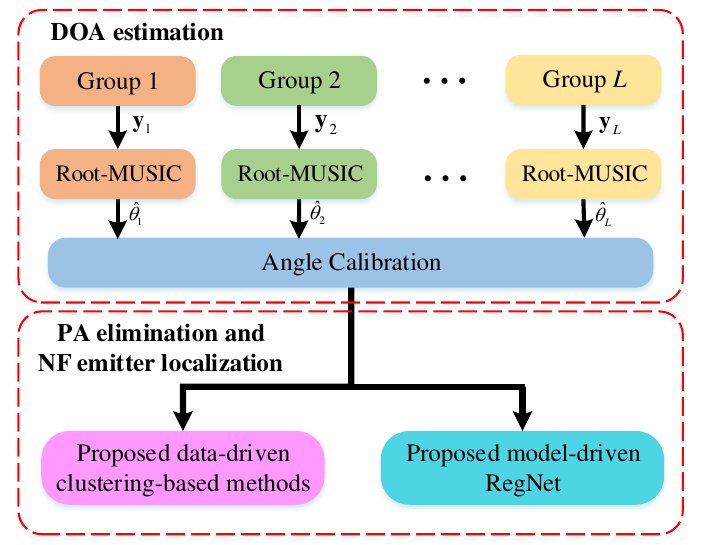}\\
	\caption{The proposed NF emitter localization framework.}\label{framework}
	\vspace{-1em}	
\end{figure}

	In order to reduce the complexity and improve the localization performance, we develop a high-performance NF emitter localization framework, as depicted in Fig. \ref{framework}. This proposed framework consists of four main steps, i.e., Root-MUSIC based DOA estimation via grouped PC-HAD structure, angle calibration, PA elimination and NF emitter localization, which will be described in the following sections.
	 
	\section{Proposed Grouped PC-HAD structure for DOA Estimation of NF Emitter}
	In this section, a grouped PC-HAD structure is proposed to transform the NF DOA estimation problem into multiple FF problems, and thus Root-MUSIC algorithm is utilized. Furthermore, to eliminate the inconsistency between these groups, an angle calibration is also developed.
	\vspace{-1em}
	\subsection{Grouped PC-HAD Structure}
	An emitter can be viewed as NF if it falls in the Fresnel region of the receive array, i.e., $[0.62(D^3/\lambda)^{1/2},2D^2/\lambda]$. The range of Fresnel region depends both on $\lambda$ and $D$. By maintaining a constant distance between emitter and receive array, increasing the signal wavelength or reducing the array aperture can effectively decrease the extent of the Fresnel region, and leading to the conversion of signal model into planar wave-based FF model. Nevertheless, in practical scenarios, the signal wavelength and array aperture are typically predetermined, so making this conversion unattainable. In this paper, we first partition the PC-HAD array into $L$ even groups, each group contains $G$ subarrays, i.e., $K=LG$, then by treating these group as independent arrays, the array aperture will be much smaller than the original receive array. Under the assumption that the following condition  
	\begin{equation}
		r>\frac{2D_g^2}{\lambda}\Rightarrow D_g<\sqrt{\frac{r\lambda}{2}},
		\label{condition}
	\end{equation}
 is met, where $D_g$ denotes the aperture of each group, it is feasible to regard the signal received within each group is planar wave. 
 	
\begin{figure}[t]
	\centering
    \includegraphics[width=0.4\textwidth]{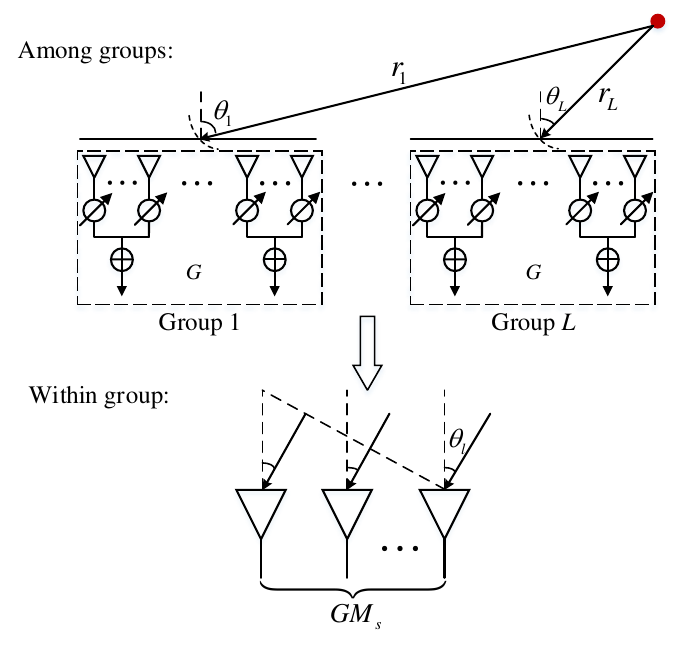}
	\caption{The proposed grouped PC-HAD structure.}\label{subarray method}
	\vspace{-1em}	
\end{figure}

As depicted in Fig. \ref{subarray method}, for a comprehensive receive array, NF signal arrives in the form of spherical wave, so the signal directions detected at reference points of distinct groups are different. However, within each group, NF signal is converted into FF, resulting in all antennas within the same group detecting signals from the same direction. Consequently, a NF signal can be simplified as $L$ distinct FF signals received by $L$ respective arrays, and the signal model is transformed to
\begin{equation}
	\mathbf{y}(t)=[\mathbf{y}_1^T(t)~\mathbf{y}_2^T(t)\cdots\mathbf{y}_L^T(t)]^T,\label{FF_model}
\end{equation}
where
\begin{equation}
	\mathbf{y}_l(t)=\mathbf{W}_l^H\tilde{\mathbf{x}}_l=\mathbf{W}_l^H\left(\mathbf{a}(\theta_l)s(t)+\mathbf{v}_l(t)\right)
\end{equation}
is the received signal model of group $l$, $\mathbf{W}_l={\rm{diag}}\left\{\mathbf{w}_{l,1},\mathbf{w}_{l,2},\cdots,\mathbf{w}_{l,G}\right\}\in \mathbb{C}^{M_g\times G}$, and
\begin{equation}
	\begin{aligned}
		\mathbf{a}(\theta_l)&=[1,e^{j\frac{2\pi}{\lambda}d\sin\theta_l},\cdots,e^{j\frac{2\pi}{\lambda}(GM_s-1)d\sin\theta_l}]^T\\
		&=\mathbf{a}_g(\theta_l)\otimes\mathbf{a}_s(\theta_l)\in\mathbb{C}^{GM_s\times 1}
	\end{aligned}\label{FF array steering vector}
\end{equation} 
is the array steering vector of group $l$, $\theta_l$ is the DOA that detected by group $l$,
$\mathbf{a}_s(\theta_l)=[1, e^{j\frac{2\pi}{\lambda}d\sin\theta_l},\cdots,e^{j\frac{2\pi}{\lambda}(M_s-1)d\sin\theta_l}]^T\in\mathbb{C}^{M_s\times 1}$, and $\mathbf{a}_g(\theta_l)=[1, e^{j\frac{2\pi}{\lambda}M_sd\sin\theta_l},\cdots,e^{j\frac{2\pi}{\lambda}(G-1)M_sd\sin\theta_l}]^T\in\mathbb{C}^{G\times 1}$ is generated by treating each subarray as an antenna and so the group can be regarded as a $G$-antennas ULA with inter-antennas spacing $M_sd$.

\vspace{-1em}
\subsection{DOA Estimation in Grouped PC-HAD}\label{NF_RMUSIC}
As the large-scale receive array is divided into $L$ small-scale groups, and the NF DOA estimation problem is transformed to $L$ FF ones, so within each group, the DOA estimation problem can be solved by low-complexity FF DOA estimation algorithm. Firstly, the covariance matrix of $\mathbf{y}_l(t)$ is given as
\begin{equation}
	\mathbf{R}_l={\rm{E}}[\mathbf{y}_l(t)\mathbf{y}_l^H(t)]=\sigma_s^2\mathbf{a}_W(\theta_l)\mathbf{a}_W^H(\theta_l)+\sigma_v^2\mathbf{I}_{G},
\end{equation}
where $\mathbf{a}_W(\theta_l)=\mathbf{W}_l^H\left(\mathbf{a}_g(\theta_l)\otimes\mathbf{a}_s(\theta_l)\right)$.
Then the MUSIC spectrum can be obtained as
\begin{equation}
	P(\theta_l)=\frac{1}{\mathbf{a}^H_W(\theta_l)\mathbf{U}_{N,l}\mathbf{U}_{N,l}^H\mathbf{a}_W(\theta_l)},\label{p_theta}
\end{equation}
where $\mathbf{U}_{N,l}\in\mathbb{C}^{G\times(G-1)}$ is corresponding to the noise subspace.
By searching the peak of the pseudo-spectrum we can obtain the DOA estimation result of each group. The essence of MUSIC algorithm is based on grid search, which requires increasing the grid density to improve accuracy, but this will also lead to a rise in algorithm complexity. Therefore, Root-MUSIC algorithm is considered as a substitution, which has the characteristics of search-free and low-complexity \cite{barabell1983improving}.

Referring to the method in \cite{shu2018low}, we set the phases of phase shifters in the group $l$ to be identical, i.e., $\mathbf{w}_{l,1}=\cdots=\mathbf{w}_{l,G}=\frac{e^{j\alpha_l}}{\sqrt{M_s}}\boldsymbol{1}_{M_s}$. Then we can get 
$\mathbf{W}_l=\mathbf{I}_G\otimes\mathbf{w}_g$,
and $\mathbf{a}_W(\theta_l)$ can be further given as
\begin{equation}
	\begin{aligned}
		\mathbf{a}_W(\theta_l)=(\mathbf{I}_G\otimes\mathbf{w}_g^H)\left(\mathbf{a}_g(\theta_l)\otimes\mathbf{a}_s(\theta_l)\right)=\mathbf{w}_{l,g}^H\mathbf{a}_s(\theta_l)\mathbf{a}_g(\theta_l),
	\end{aligned}
\end{equation}
where
\begin{equation}
	\begin{aligned}
		\mathbf{w}_{l,g}^H\mathbf{a}_s(\theta_l)&=\frac{e^{j\alpha_l}}{\sqrt{M_s}}\sum_{m_s=1}^{M_s}e^{j\frac{2\pi}{\lambda}(m_s-1)d\sin\theta_l}\\
		&=\frac{e^{j\alpha_l}}{\sqrt{M_s}}\cdot\frac{1-e^{j\frac{2\pi}{\lambda}M_sd\sin\theta_l}}{1-e^{j\frac{2\pi}{\lambda}d\sin\theta_l}}.
	\end{aligned}
\end{equation}
So it makes sense that the new expression of $P(\theta_l)$ is
\begin{equation}
	P(\theta_l)=\frac{1}{\lVert\mathbf{w}_{l,g}^H\mathbf{a}_s(\theta_l)\rVert^2\mathbf{a}^H_g(\theta_l)\mathbf{U}_N\mathbf{U}_N^H\mathbf{a}_g(\theta_l)},
\end{equation}
and let $\mathbf{C}=\mathbf{U}_N\mathbf{U}_N^H$, we can obtain
\begin{equation}
	\begin{aligned}
		&P^{-1}(\theta_l)=\frac{2-e^{j\frac{2\pi}{\lambda}M_sd\sin\theta_l}-e^{-j\frac{2\pi}{\lambda}M_sd\sin\theta_l}}{2-e^{j\frac{2\pi}{\lambda}d\sin\theta_l}-e^{-j\frac{2\pi}{\lambda}d\sin\theta_l}}\cdot\\
		&\sum_{g_1=1}^{G}\sum_{g_2=1}^{G}e^{-j\frac{2\pi}{\lambda}(g_1-1)M_sd\sin\theta_l}\mathbf{C}_{g_1,g_2}e^{j\frac{2\pi}{\lambda}(g_2-1)M_sd\sin\theta_l},\label{P_inverse}
	\end{aligned}
\end{equation}
where $\mathbf{C}_{g_1,g_2}$ denotes the element at the $g_2$th column of the $g_1$th row of $\mathbf{C}$. Then following the procedure of Root-MUSIC algorithm, we define  $z_l=e^{j\frac{2\pi}{\lambda}M_sd\sin\theta_l}$ and the equivalent polynomial is expressed as
\begin{equation}
	\begin{aligned}
	f(z_l)=
	\frac{2-z_l-z_l^{-1}}{2-z_l^{\frac{1}{M_s}}-z_l^{-\frac{1}{M_s}}}
	\sum_{g_1=1}^{G}\sum_{g_2=1}^{G}z_l^{-(g_1-1)}\mathbf{C}_{g_1,g_2}z_l^{g_2-1}.
	\end{aligned}
\end{equation}

Therefore, searching the peak of (\ref{p_theta}) is equivalent to evaluating the root of polynomial $f(z_l)$ that be closest to the unit circle, which is denoted by $\hat{z}_l$, 
then the signal direction $\theta_l$ can be estimated as
\begin{equation}
	\hat{\theta}_l=\arcsin\left(\frac{\lambda\arg(\hat{z}_l)}{2\pi M_sd}\right).\label{ambiguity_theta}
\end{equation}
However, as each group in the grouped PC-HAD structure is viewed as a $G$-antennas ULA and with inter-antennas spacing $M_s\lambda/2$, this will cause an ambiguity with $M_s$ possible results in phase extraction. So in order to solve this ambiguity, we first rewritten (\ref{ambiguity_theta}) as
\begin{equation}
	\hat{\theta}_{l,i}=\arcsin\left(\frac{\lambda(\arg(\hat{z}_l)+2\pi i)}{2\pi M_sd}\right),\label{estimated_theta}
\end{equation}
where $i\in\{0,1,\cdots,M_s-1\}$ is the ambiguity coefficient. Then a solution set contained $M_s$ elements can be constructed for group $l$:
\begin{equation}
	\hat{\Theta}_l=\left\{\hat{\theta}_{l,0},\hat{\theta}_{l,1},\cdots,\hat{\theta}_{l,M_s-1}\right\}\in\mathbb{R}^{M_s},\label{hat_Theta_l}
\end{equation}
where $\hat{\Theta}_l$ is called initial set in this work, $\hat{\theta}_{l,i}$ stands for the initial solution of group $l$ with ambiguity coefficient $i$. It is obviously that there is only one true solution in $\hat{\Theta}_l$, which is denoted by $\hat{\theta}_{l,true}=\theta_l$, and the rest solutions are false solutions.
So the problem here is how to eliminate the false solutions, and thus obtaining the true DOA.  
\vspace{-1em}
\subsection{Angle Calibration for Grouped PC-HAD}
\begin{figure}[!t]
	\centering
	\includegraphics[width=0.35\textwidth]{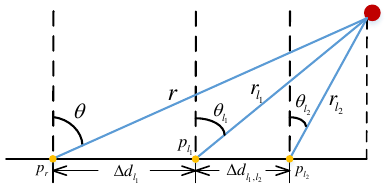}\\
	\caption{The geometric relationship between the reference points of various groups and the entire array in the NF model.}\label{angle}
	\vspace{-1em}	
\end{figure}

Since the DOAs measured by $L$ groups are different from each other for the NF emitter, it is necessary to calibrate them to a common point to obtain the final DOA estimation result.
As shown in Fig. \ref{angle}, $p_r$ is a reference point of the whole receive array, $p_{l_1}$ and $p_{l_2}$ denote the reference points of group $l_1$ and $l_2$, where $1\leq l_1<l_2\leq L$. Then the position of the emitter relative to these three reference points can be respectively represented as $(\theta,r)$, $(\theta_{l_1},r_{l_1})$ and $(\theta_{l_2},r_{l_2})$, where $\theta_{l_1}$ and $\theta_{l_2}$ that can be obtained by (\ref{estimated_theta}). From the Fig. \ref{angle}, we can observe that $(\theta,r)$ can be expressed by $(\theta_{l_1},r_{l_1})$ and $(\theta_{l_2},r_{l_2})$ based on their geometrical relationship. Because there are four unknown parameters: $\theta$, $r$, $r_{l_1}$ and $r_{l_2}$, at least four independent equations are needed to solve this problem. In this way, the following equations are established: 
\begin{subequations}
	\begin{align}
		&r_{l_1}\cos\theta_{l_1}=r_{l_2}\cos\theta_{l_2},\label{a}\\
		&r_{l_1}\sin\theta_{l_1}=r_{l_2}\sin\theta_{l_2}+\Delta d_{l_1,l_2},\label{b}\\
		&r\cos\theta=r_{l_2}\cos\theta_{l_2},\label{c}\\
		&r\sin\theta=r_{l_2}\sin\theta_{l_2}+\Delta d_{l_2},\label{d}
	\end{align}
\end{subequations}
where $\Delta d_{l_1,l_2}=(l_2-l_1)GM_sd$ denotes the distance between $p_{l_1}$ and $p_{l_2}$, $\Delta d_{l_2}=\Delta d_{l_1}+\Delta d_{l_1,l_2}$ is the distance between $p_r$ and $p_{l_2}$.
Firstly, substituting (\ref{a}) into (\ref{b}), we can get the expression of $r_{l_2}$ as
\begin{equation}
	r_{l_2}=\frac{\Delta d_{l_1,l_2}\cos\theta_{l_1}}{\sin\theta_{l_1}\cos\theta_{l_2}-\cos\theta_{l_1}\sin\theta_{l_2}}.\label{r_l2}
\end{equation}
From (\ref{c}), we can know $r=r_{l_2}\cos\theta_{l_2}/\cos\theta$. Substituting (\ref{r_l2}) into (\ref{d}), the following equation is generated
\begin{equation}
	\tan\theta=\frac{\Delta d_{l_2}\tan\theta_{l_1}-\Delta d_{l_1}\tan\theta_{l_2}}{\Delta d_{l_1,l_2}}.
\end{equation}

By substituting the DOAs estimated by any two groups, the calibrated $\theta$ can also be obtained as
\begin{equation}
	\hat{\theta}_{l_1,l_2,i}=\arctan\left(\frac{\Delta d_{l_2}\tan\hat{\theta}_{l_1,i}-\Delta d_{l_1}\tan\hat{\theta}_{l_2,i}}{\Delta d_{l_1,l_2}}\right),\label{calibrated theta}
\end{equation}
where $\hat{\theta}_{l_1,l_2,i}$ represents the calibrated angle obtained by combining group ${l_1}$ and group ${l_2}$. As the reference point of a receive array is usually located at the first antenna, then $p_r$ and $p_{l_1}$ are overlapped when $l_1=1$, and from Fig. \ref{angle} we know $\Delta d_1=0$ and $\Delta d_{1,l_2}=\Delta d_{l_2}$ in this case. Therefore, based on (\ref{calibrated theta}) we can get
\begin{equation}
	\begin{aligned}
		\hat{\theta}_{1,l_2,i}=\arctan(\tan\hat{\theta}_{1,i})=\begin{cases}
			\hat{\theta}_{1,i}, & -\frac{\pi}{2}\leq\hat{\theta}_1\leq\frac{\pi}{2}\\
			\hat{\theta}_{1,i}-\pi, & \hat{\theta}_1>\frac{\pi}{2}\\
			\hat{\theta}_{1,i}+\pi, & \hat{\theta}_1<-\frac{\pi}{2}
		\end{cases},\label{equation}
	\end{aligned}
\end{equation}
where $2\leq l_2\leq L$. Because the range of $\theta$ is restricted in $[-\pi/2,\pi/2]$, if $\hat{\theta}_{1,l_2,i}\neq \hat{\theta}_{1,i}$, the corresponding $\hat{\theta}_{1,i}$ is a definite false solution.

Since the ambiguity coefficient has not been determined, the calibrated set also contains one true solution and $M_s-1$ false solutions, just like (\ref{hat_Theta_l}):
\begin{equation}
	\hat{\Theta}_{l_1,l_2}=\left\{\hat{\theta}_{l_1,l_2,0},\hat{\theta}_{l_1,l_2,1},\cdots,\hat{\theta}_{l_1,l_2,M_s-1}\right\}\in\mathbb{R}^{M_s},\label{hat_Theta_l1_l2}
\end{equation}
and the true solution of this set is denoted by $\hat{\theta}_{l_1,l_2,i_t}$.
In noiseless situation, the true solutions of different sets are identical, i.e., $\hat{\theta}_{1,2,i_t}=\cdots=\hat{\theta}_{L-1,L,i_t}=\theta$,
where $i_t$ denotes the true ambiguity coefficient.
And the false solutions are usually different from each other.
Then combining all $L$ groups, we can get a calibrated angle set
\begin{equation}
	\hat{\Theta}=\hat{\Theta}_{1,2}\cup\hat{\Theta}_{1,3}\cup\cdots\cup\hat{\Theta}_{L-1,L}\in\mathbb{R}^{NM_s}.
\end{equation}

As the range parameter $r$ can be obtained by (\ref{c}), then the calibrated range set is also given as 
\begin{equation}
	\hat{\mathcal{R}}=\hat{\mathcal{R}}_{1,2}\cup\hat{\mathcal{R}}_{1,3}\cup\cdots\cup\hat{\mathcal{R}}_{L-1,L}\in\mathbb{R}^{NM_s},\label{distance set}
\end{equation}
where $\hat{\mathcal{R}}_{l_1,l_2}=\left\{\hat{r}_{l_1,l_2,0},\hat{r}_{l_1,l_2,1},\cdots,\hat{r}_{l_1,l_2,M_s-1}\right\}$ and
\begin{equation}
	\hat{r}_{l_1,l_2,i}=\frac{\hat{r}_{l_2,i}\cos\hat{\theta}_{l_2,i}}{\cos\hat{\theta}_{l_1,l_2,i}}.
\end{equation}
The true solutions of subsets that compose $\hat{\mathcal{R}}$ also satisfy $\hat{r}_{1,2,i_t}=\cdots=\hat{r}_{L-1,L,i_t}=r$.

By matching the elements in $\hat{\Theta}$ and $\hat{\mathcal{R}}$ one by one, we can get a candidate position set for the NF emitter 
\begin{equation}
	\Omega=\left\{\boldsymbol{\omega}_{l_1,l_2,i }\big|1\leq l_1<l_2\leq L,~0\leq  i\leq M_s-1\right\},\label{omega}
\end{equation}
where $\omega_{l_1,l_2,i}=(\hat{\theta}_{l_1,l_2,i},\hat{r}_{l_1,l_2,i})$ and $|\Omega|=NM_s$. Based on different ambiguity coefficients, $\Omega$ can be naturally divided into $M_s$ clusters:
\begin{equation}
	\Omega=\Omega_0\cup\Omega_1\cup\cdots\cup\Omega_{M_s-1},
\end{equation}
where $\Omega_i=\left\{\boldsymbol{\omega}_{l_1,l_2,i }\big|1\leq l_1<l_2\leq L\right\}$,
and $|\Omega_i|=N$. Among theses $M_s$ clusters, there are one true solution cluster and $M_s-1$ false solution clusters. The true solution cluster is given as
\begin{equation}
	\Omega_{true}=\left\{\boldsymbol{\omega}_{l_1,l_2,i_t }\big|1\leq l_1<l_2\leq L\right\},
\end{equation}
where $\boldsymbol{\omega}_{l_1,l_2,i_t}=(\hat{\theta}_{l_1,l_2,i_t},\hat{r}_{l_1,l_2,i_t})$. And we have $\boldsymbol{\omega}_{1,2,i_t}=\cdots=\boldsymbol{\omega}_{L-1,L,i_t}$ under noiseless condition.

\vspace{-0.5em}
\section{Proposed Low-Complexity Clustering-based Methods for NF Localization}
In this section, to eliminate the false solutions caused by PA and obtain the true solution cluster, two clustering-methods are developed based on the data distribution features of the samples in the candidate position set. 
\vspace{-1em}
\subsection{Minimum Sample Distance Clustering}
As the position of NF emitter can not be calculated before the ambiguity coefficient is determined, we have to find out the special characters of the true solutions to discriminate them from the false solutions in $\Omega$.
In Fig. \ref{theta-r cluster}, we plot the samples of $\Omega$ in a coordinate with angle as $x$-axis and distance as $y$-axis in the noise-free situation, the actual position of the emitter relative to the receive array is $(60^{\circ},50m)$. A cluster in this diagram is corresponding to an $\Omega_i$, 
the true ambiguity coefficient is $i_{true}=2$ in this case, and thus the samples of cluster 2 are gathered together, while the samples of other clusters are scattered especially in the distance dimension.  

Based on the distribution characters of the calibrated points in the $(\theta,r)$-diagram and taking the impact of noise into consideration, the true solution set $\Omega_{true}$ can be distinguished via finding the most concentration one among all the $M_s$ clusters.
Here the scatter degree of a cluster is defined by the summation of the distance between any two samples in this cluster, and the squared Euclidean distance (SED) is employed as distance measure. Then we can get the distance between two samples as
\begin{equation}
		d_{n_1,n_2}^i=\lVert\boldsymbol{\omega}_{n_1,i}-\boldsymbol{\omega}_{n_2,i}\rVert^2\label{d_i}
\end{equation}
where $1\leq n_1<n_2\leq N$ and $N={L\choose2}$. Here $\boldsymbol{\omega}_{l_1,l_2,i}$ is simplified to $\boldsymbol{\omega}_{n,i}$ for convenient expression. And the scatter degree of cluster $i$ is defined as
\begin{equation}
	\bar{d}_i=\sum_{n_1=1}^{N-1}\sum_{n_2=n_1+1}^{N}d_{n_1,n_2}^i.
\end{equation}
It's obviously that the smaller value of $\bar{d}_i$ represents the corresponding cluster has high concentration,
so the optimization problem is given as
\begin{equation}
	\hat{i}=\arg\min_i \bar{d}_i.
\end{equation}
Then the final estimation results of $\theta$ and $r$ can also be given as
\begin{equation}
	\begin{aligned}
		\hat{\theta}=\frac{1}{N}\sum_{n=1}^{N}\hat{\theta}_{n,\hat{i}},\quad\hat{r}=\frac{1}{N}\sum_{n=1}^{N}\hat{r}_{n,\hat{i}},
	\end{aligned}
\end{equation}
and the estimated NF emitter position is $(\hat{\theta},\hat{r})$.

\vspace{-1em}
\subsection{RSD-ASD-based Density Clustering}
\begin{figure}[!t]
	\centering
	\includegraphics[width=0.35\textwidth]{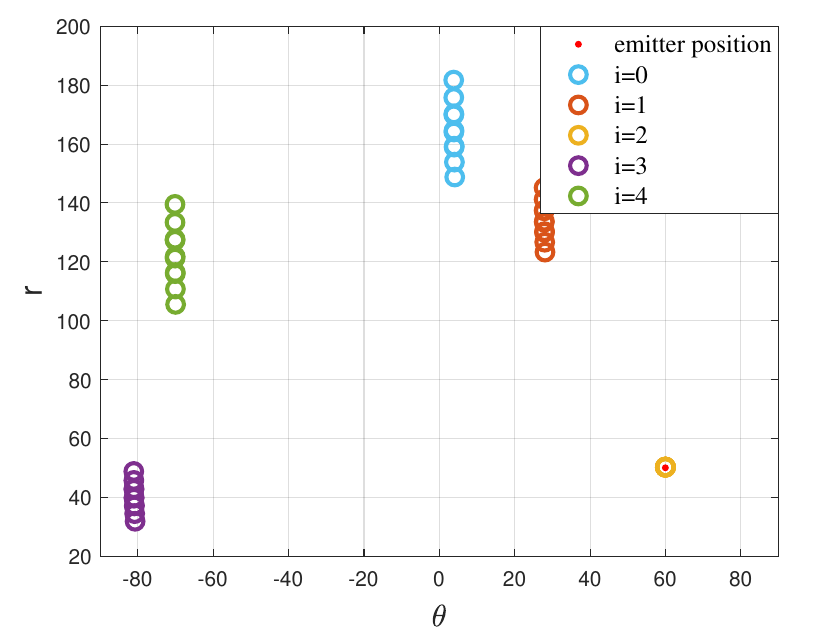}
	\caption{The calibrated candidate emitter positions discriminated by ambiguity coefficient in the $(\theta,r)$-plane.}\label{theta-r cluster}
	\vspace{-1em}	
\end{figure}

\begin{figure*}[ht]
	\centering
	\includegraphics[width=0.8\linewidth]{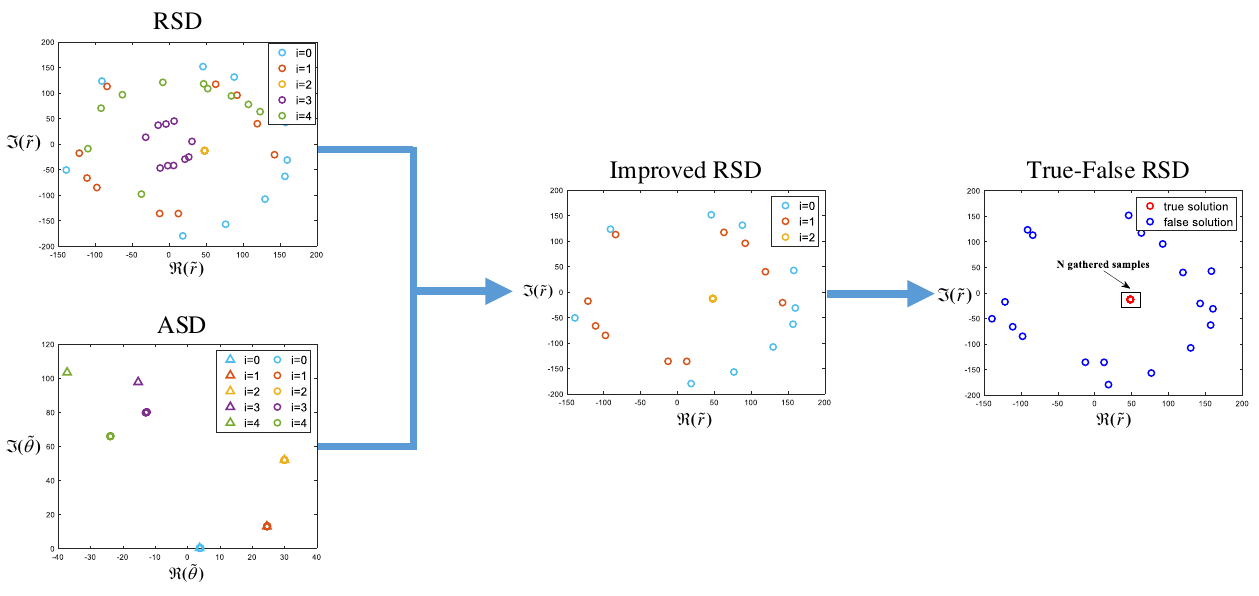}
	\caption{The proposed clustering method for NF emitter localization by combining RSD and ASD.}\label{theta-r combinition}
	\vspace{-1em}	
\end{figure*}

In Fig. \ref{theta-r cluster}, the candidate emitter positions are plotted in the $(\theta,r)$ diagram, and the distribution characteristics of the calibrated points are extracted by combining the angle and range dimensions. But we can also find that these points have different distributions in angle and range dimensions. Therefore, in order to further explore the distribution characteristics of different dimensions and improve the emitter localization accuracy, we decide to construct the angle scatter diagram (ASD) and range scatter diagram (RSD) respectively as shown in Fig. \ref{theta-r combinition}. As the elements of $\hat{\Theta}$ and $\hat{\mathcal{R}}$ just have one dimension, so we need to make a transformation that allow them to be scattered in a 2D plane. Then we first get the polarization forms of $\theta$ and $r$:
\begin{equation}
	\begin{aligned}
		\tilde{\theta}=\theta e^{j\theta},~~\tilde{r}=r e^{jr},\label{polarization}
	\end{aligned}
\end{equation}
and by extracting their real and image parts respectively, the real numbers are transformed to two-dimensional vectors, i.e., $(\Re(\tilde{\theta}),\Im(\tilde{\theta}))$ and $(\Re(\tilde{r}),\Im(\tilde{r}))$. In the following content, we will introduce how to construct RSD and ASD based on this transformation, and thus achieving high-accuracy localization of NF emitter.

Firstly, we perform the transformation on the calibrated range set $\hat{\mathcal{R}}$ and get 
\begin{equation}
	\tilde{\mathcal{R}}=\left\{\tilde{\boldsymbol{\omega}}^r_{l_1,l_2,i}\big|1\leq l_1<l_2\leq L,0\leq i\leq M_s-1\right\},
\label{polarization distance set}
\end{equation}
where $|\tilde{\mathcal{R}}|=NM_s$,  $\tilde{\boldsymbol{\omega}}^r_{l_1,l_2,i}=(\Re(\tilde{r}_{l_1,l_2,i}),\Im(\tilde{r}_{l_1,l_2,i}))$ and $\tilde{r}_{l_1,l_2,i}=\hat{r}_{l_1,l_2,i} e^{j\hat{r}_{l_1,l_2,i}}$. $\tilde{\mathcal{R}}$ can also be divided into $M_s$ subsets based on different ambiguity coefficients as $\tilde{\mathcal{R}}=\tilde{\mathcal{R}}_{0}\cup\tilde{\mathcal{R}}_{1}\cup\cdots\cup\tilde{\mathcal{R}}_{M_s-1}$, and the true solution set therein is $\tilde{\mathcal{R}}_{true}=\left\{\tilde{\boldsymbol{\omega}}^r_{l_1,l_2,i_t}\big|1\leq l_1<l_2\leq L\right\}$,
where $\tilde{\boldsymbol{\omega}}^r_{1,2,i_t}=\cdots=\tilde{\boldsymbol{\omega}}^r_{L-1,L,i_t}$.
Compared to Fig.\ref{theta-r cluster}, the false solutions in the proposed RSD are more scattered, and the true solutions are still inseparable, so RSD is more suitable for the application of high-performance clustering methods. 

In order to weaken the impact of false solutions on the final localization result, based on (\ref{equation}) and the related conclusions, we construct a new candidate angle set as
\begin{equation}
	\hat{\mathcal{A}}=\hat{\Theta}_1\cup\hat{\Theta}_{1,2}\cup\hat{\Theta}_{1,3}\cup\cdots\cup\hat{\Theta}_{1,L}\in\mathbb{R}^{LM_s},
\end{equation} 
where $\hat{\Theta}_1$ is defined in (\ref{hat_Theta_l}), and $\hat{\Theta}_{1,2}$ to $\hat{\Theta}_{1,L}$ are defined in (\ref{hat_Theta_l1_l2}). Then the polarization forms of the elements in $\hat{\mathcal{A}}$ can be obtained and the set is transformed to
\begin{equation}
	\begin{aligned}
		\tilde{\mathcal{A}}=&\left\{\tilde{\boldsymbol{\omega}}^{\theta}_{1,i}\big|0\leq i\leq M_s-1\right\}\cup\\
		&\left\{\tilde{\boldsymbol{\omega}}^{\theta}_{1,l,i}\big|2\leq l\leq L,0\leq i\leq M_s-1\right\},
	\end{aligned}
	\label{polarization angle set}
\end{equation}
where $|\tilde{\mathcal{A}}|=LM_s$,  $\tilde{\boldsymbol{\omega}}^{\theta}_{1,i}=(\Re(\tilde{\theta}_{1,i}),\Im(\tilde{\theta}_{1,i}))$ and $\tilde{\theta}_{1,i}=\hat{\theta}_{1,i} e^{j\hat{\theta}_{1,i}}$, $\tilde{\omega}^{\theta}_{1,l,i}=(\Re(\tilde{\theta}_{1,l,i}),\Im(\tilde{\theta}_{1,l,i}))$ and $\tilde{\theta}_{1,l,i}=\hat{\theta}_{1,l,i} e^{j\hat{\theta}_{1,l,i}}$. So the ASD in Fig.\ref{theta-r combinition} is constructed by the points of $\tilde{\mathcal{A}}$, and to be distinguished easily, $\tilde{\boldsymbol{\omega}}^{\theta}_{1,i}$ is plotted by triangular symbol. From (\ref{equation}), we can know if $i$ is the true ambiguity coefficient, then $\tilde{\boldsymbol{\omega}}^{\theta}_{1,i}=\tilde{\boldsymbol{\omega}}^{\theta}_{1,2,i}=\cdots=\tilde{\boldsymbol{\omega}}^{\theta}_{1,L,i}$ is the necessary and insufficient condition. But it's obviously that the clusters of $i=3$ and $i=4$ in the ASD can't satisfy this condition, so they can be labeled as false ambiguity coefficients directly and the corresponding angles are false solutions. 

By combining RSD and ASD, we can eliminate a part of false solutions clusters corresponding to $i$ which have been labeled as false ambiguity coefficients by ASD, then the improved range solution set is obtained as
\begin{equation}
	\tilde{\mathcal{R}}_{imp}=\tilde{\mathcal{R}}_{imp,1}\cup\cdots\cup\tilde{\mathcal{R}}_{imp,M_{imp}},
\end{equation}
where $M_{imp}\leq M_s$ denotes the number of remaining clusters after the false solution clusters elimination and $|\tilde{\mathcal{R}}_{imp}|=NM_{imp}$,
thereby generating the improved RSD. The step 3 of algorithm \ref{DSD-ASD DBSCAN} describes how to eliminate the false solutions based on $\tilde{\mathcal{A}}$ in the actual operations, where $\varepsilon$ is a threshold which decreases with the increasing of $\rm{SNR}$ and $\varepsilon=0$ under the noise-free condition.
Compared to the initial RSD, the improved RSD has less false solutions, and thus the true solutions are easier to be identified.
Then since the true solutions gather together, density-based clustering methods can be considered for inferring the true solutions.

As the most representative algorithm for density-based clustering, the core idea of DBSCAN \cite{ester1996density} is to measure the space density of a sample point by the number of points within its neighborhood. Therefore, there are two parameters requiring to be given before clustering: $\epsilon$, the radius of neighborhood, and $MinPts$ denotes the minimum number of sample points that contained in the $\epsilon$-neighborhood. 
Next we will introduce how to determine the neighborhood parameter based on our problem. Since the true solution set $\tilde{\mathcal{R}}_{true}$ has $N$ points, we can get $MinPts=N$. Then in order to obtain the neighborhood range $\epsilon$, first we calculate the distances between all sample points within each subset of $\tilde{\mathcal{R}}_{imp}$, and then find the maximum value among them 
\begin{equation}
	d_{m,max}=\max\left\{\lVert\boldsymbol{\omega}_{m,n_1}^r-\boldsymbol{\omega}_{m,n_2}^r\rVert^2|1\leq n_1<n_2\leq N\right\},
\end{equation}
where $\boldsymbol{\omega}_{m,n}^r$ denotes the $n$-th sample of $\tilde{\mathcal{R}}_{imp,m}$ and $1\leq m\leq M_{imp}$. Due to the elements of true solution set have the tightest connections, so the value of $\epsilon$ is set as the minimum $d_{m,max}$, i.e.,
\begin{equation}
	\epsilon=\min\left\{d_{m,max}|1\leq m\leq M_{imp}\right\}.
\end{equation}
Since the required parameters are determined, we can perform clustering on sample set $\tilde{\mathcal{R}}_{imp}$ via DBSCAN, and the clustering procedure can be summarized as: (1) find out the samples that have at least $N$ points in their $\epsilon$-neighborhoods and label their as core objects, (2) all the core objects that are density-connected and the points in their neighborhoods are divided into the same clusters. Then the clustering result can be expressed by
\begin{equation}
	\left\{\mathcal{C}_{n_c}|1\leq n_c\leq N_c\right\}={\rm{DBSCAN}}\left(\tilde{\mathcal{R}}_{imp},(\epsilon,N)\right)\label{dbscan}
\end{equation}
where $N_c$ denotes the number of clusters. Next, we seek out the cluster $\mathcal{C}_{n_c}=\mathcal{C}_{max}$ that has the most points, i.e., $|\mathcal{C}_{max}|=\max\{|\mathcal{C}_{1}|,|\mathcal{C}_{2}|,\cdots,|\mathcal{C}_{N_c}|\}$. Since the task is to find the $N$ true solutions, if $|\mathcal{C}_{max}|=N$ we can get $\tilde{\mathcal{R}}_{true}=\mathcal{C}_{max}$, but if $|\mathcal{C}_{max}|>N$, it needs to be further divided until $|\mathcal{C}_{max}|=N$ is achieved. In this process, the neighborhood radius $\epsilon$ should be reduced in each iteration, and the renewed $\epsilon$ is given as
\begin{equation}
	\epsilon_j=\eta_j\epsilon_{j-1},
\end{equation} 
where $j$ denotes the iteration number and $\eta_j<1$ is an adjustable coefficient.

With the obtained true point set $\tilde{\mathcal{R}}_{true}=\{\tilde{\omega}_{true,1},\tilde{\omega}_{true,2}\cdots\tilde{\omega}_{true,N}\}$, where $\tilde{\omega}_{true,n}=(\Re(\tilde{r}_{true,n}),\Im(\tilde{r}_{true,n}))$, we can get the true range solution set $\hat{\mathcal{R}}_{true}=\{\hat{r}_{true,1},\hat{r}_{true,2},\cdots,\hat{r}_{true,N}\}$ recovered from $\tilde{\mathcal{R}}_{true}$ based on (\ref{polarization}), where $\hat{r}_{true,n}=|\tilde{\omega}_{true,n}|$. Then the range estimation result is given as
\begin{equation}
	\hat{r}=\frac{1}{N}\sum_{n=1}^{N}\hat{r}_{true,n}.\label{hat_r_dbscan}
\end{equation}
Since the initial sample set $\tilde{\mathcal{R}}_{imp}$ comes from the calibrated distance set (\ref{distance set}) and $\tilde{\mathcal{R}}_{true}\subseteq \tilde{\mathcal{R}}_{imp}$, then we can get $\hat{\mathcal{R}}_{true}\subset\hat{\mathcal{R}}$. So we can always find a corresponding angle $\hat{\theta}_{true,n}$ for $\hat{r}_{true,n}$ from $\Omega$, and the angle estimation result can also be given as 
\begin{equation}
	\hat{\theta}=\frac{1}{N}\sum_{n=1}^{N}\hat{\theta}_{true,n}.\label{hat_theta_dbscan}
\end{equation}

\begin{algorithm}[tb]
	\caption{RSD-ASD-DBSCAN}\label{DSD-ASD DBSCAN}
	\begin{algorithmic}
		\Require $\hat{\Theta}_{1}$: initial set obtained by group 1, $\hat{\Theta}$: calibrated angle set, $\hat{\mathcal{R}}$: calibrated range set;
		\State 1: Perform polarization on $\hat{\mathcal{R}}$, and extract its real and imaginary parts respectively to construct $\tilde{\mathcal{R}}$, then RSD is generated based on $\tilde{\mathcal{R}}$;
		\State 2: Obtain $\hat{\mathcal{A}}$ by combining $\hat{\Theta}_{1}$ and $\hat{\Theta}$, same as step 1, $\hat{\mathcal{A}}$ can be transformed to $\tilde{\mathcal{A}}$ and ASD is generated;
		\State 3: Use principle (\ref{equation}) on ASD, a part of false ambiguity coefficients are distinguished, and by eliminating the corresponding false solution clusters, the improved range set $\tilde{\mathcal{R}}_{imp}$ can be obtained. Given $\tilde{\mathcal{R}}_{imp}=\tilde{\mathcal{R}}$, this procedure is given as:
		\State~\quad 3.1: \textbf{for} {$i=0,1,\cdots,M_s-1$} \textbf{do}
		\State~\quad 3.2: \quad~ \textbf{if}  {$\sum_{l=2}^{L}\lVert\tilde{\omega}^{\theta}_{1,l,i}-\tilde{\omega}^{\theta}_{1,i}\rVert^2>\varepsilon$}
		\State~\quad 3.3: \quad\quad~ $\tilde{\mathcal{R}}_{imp}=\tilde{\mathcal{R}}_{imp}/\tilde{\mathcal{R}}_i$;
		\State~\quad 3.4: \quad~ \textbf{else}
		\State~\quad 3.5: \quad\quad~ $\tilde{\mathcal{R}}_{imp}=\tilde{\mathcal{R}}_{imp}$;
		\State~\quad 3.6: \quad~ \textbf{end if} 
		\State~\quad 3.7: \textbf{end for} 
		\State 4: Given the initial neighborhood parameters $(\epsilon_0,N)$ and implement DBSCAN on $\tilde{\mathcal{R}}_{imp}$ to find the cluster that has the highest density, repeat this operation until the size of the obtained cluster is $N$. The concrete procedure is given as:
		\State~\quad 4.1: \textbf{while} {$|\mathcal{C}_{max,j-1}|>N$} \textbf{do}
		\State~\quad 4.2: \quad $\epsilon_{j}=\eta_j\epsilon_{j-1}$;
		\State~\quad 4.3: \quad implement (\ref{dbscan}) and find out $\mathcal{C}_{max,j}$;
		\State~\quad 4.4: \quad $\tilde{\mathcal{R}}_{imp}=\mathcal{C}_{max,j}$;
		\State~\quad 4.5: \quad $j=j+1$;
		\State~\quad 4.6: \textbf{end while}
		\State 5: Obtain the true range cluster and infer the corresponding angle cluster from (\ref{omega});
		\State 6: Calculate the range and angle estimation results based on (\ref{hat_r_dbscan}) and (\ref{hat_theta_dbscan});
		\Ensure $\hat{r}$, $\hat{\theta}$.
	\end{algorithmic}
\end{algorithm}

\vspace{-1em}
\section{Proposed Regression Network for High-resolution NF Localization} 
In this section, a regression network is proposed for achieving higher localization accuracy by combining the functions of false solution elimination and true solution fusion.
\vspace{-1em}
\subsection{Network Construction}
By combining the initial sets of all the groups, we can get 
\begin{equation}
	\hat{\Theta}=\hat{\Theta}_1\cup\hat{\Theta}_2\cup\cdots\cup\hat{\Theta}_L\in\mathbb{R}^{LM_s},
\end{equation}
from the generation mechanism of the phase ambiguity we can know there are $L(M_s-1)$ false solutions in $\hat{\Theta}$, so the task of phase ambiguity elimination is to find out the $L$ true solution from all $LM_s$ solutions. As this process is nonlinear, then it can be simply expressed as 
\begin{equation}
	\hat{\Theta}_{true}=f_{nl}(\hat{\Theta}),\label{nonlinear mapping}
\end{equation}
where $\hat{\Theta}_{true}=\{\hat{\theta}_{1,true},\cdots,\hat{\theta}_{L,true}\}$ and $f_{nl}(\cdot)$ denotes an nonlinear mapping relationship. By observing (\ref{nonlinear mapping}) we can find, since we only know the input and output, and don't know the concrete mapping relationship, thus this problem can't be transformed and solved by linear optimization algorithms. Then machine learning-based methods are usually adopted to solve problems like this. Depending on the specific situations and goals to be achieved, common methods include kernel methods \cite{cortes1995support}, neural networks \cite{bishop1995neural}\cite{rumelhart1986learning}, non-parametric methods \cite{silverman2018density}, etc. Among them, neural networks have the best performance for they can learn complex nonlinear mapping relationships through multiple layers of nonlinear transformations. Therefore, we will design a multi-layer neural network (MLNN) for solving the problem (\ref{nonlinear mapping}).

\begin{figure}[t]
	\centering
	\includegraphics[width=0.45\textwidth]{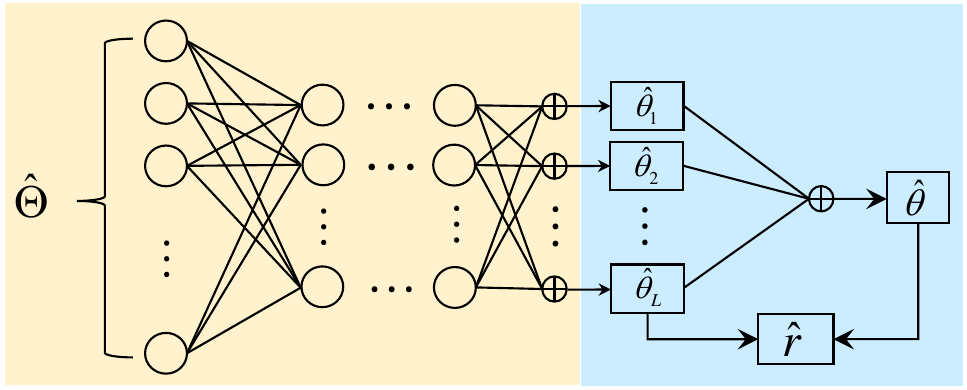}
	\caption{The proposed RegNet for NF localization, where the yellow part is a MLNN that performs false solutions elimination and the blue part is a perceptron for localization.}\label{regression network}	
	\vspace{-1em}
\end{figure}

As shown in the yellow part of Fig.\ref{regression network}, the input data is $\hat{\Theta}$, so the input layer of this MLNN contains $LM_s$ neurons, and the output layer has $L$ neurons for the output is $\hat{\Theta}_{true}$. Supposing the MLNN has $H$ hidden layers, then (\ref{nonlinear mapping}) is transformed to
\begin{equation}
	\hat{\boldsymbol{\Theta}}_{true}=f^{(\rm{out})}(f^{(H)}(\cdots f^{(1)}(\mathbf{W}^{(\rm{in})}\hat{\boldsymbol{\Theta}}+\mathbf{b}^{(\rm{in})}))),
\end{equation}
where $\hat{\boldsymbol{\Theta}}_{true}=[\hat{\theta}_{1,true},\cdots,\hat{\theta}_{L,true}]^T\in\mathbb{R}^{L\times 1}$ and $\hat{\boldsymbol{\Theta}}=[\hat{\theta}_{1,0},\cdots,\hat{\theta}_{1,M_s-1},\cdots,\hat{\theta}_{L,0},\cdots,\hat{\theta}_{L,M_s-1}]^T\in\mathbb{R}^{LM_s\times 1}$ are vectors that including the elements of sets $\hat{\Theta}_{true}$ and $\hat{{\Theta}}$, $\mathbf{W}^{(\rm{in})}$ and $\mathbf{b}^{(\rm{in})}$ are weight and bias vectors of input layer, $f^{(h)}(\cdot)$ represents the nonlinear transformation of the $h$th hidden layer. As the nonlinear modeling ability of neural networks comes from the activation functions, so here rectified linear unit (ReLU) is adopted as activation functions for the hidden layers. Finally, the expected output vector $\hat{\boldsymbol{\Theta}}_{true}$ contains $L$ specific angles, so this MLNN is designed for achieving a regression task, then the output layer can adopt linear activation function.

Since the emitter is an NF emitter, the true solutions of $L$ groups are different from each other, i.e., $\hat{\theta}_{1,true}\neq\hat{\theta}_{2,true}\neq\cdots\neq\hat{\theta}_{L,true}$. Therefore, $\hat{\boldsymbol{\Theta}}_{true}$ isn't the final DOA estimation result, we have to design another method to fuse these true solutions. As there must be a linear transformation between two numbers, the fusion function can be given as the linear combination of the $L$ true angle solutions:
\begin{equation}
	\hat{\theta}_{true}=\sum_{l=1}^{L}\beta_lf_l(\hat{\theta}_{l,true}),\label{linear combination}
\end{equation}
where $\beta_l$ is the combination coefficient and $f_l(\cdot)$ denotes a linear transformation. Due to the optimal $\beta_l$ and transformation relationship are unknown, we can design a perceptron like the blue part of Fig. \ref{regression network} to solve this problem, and thus (\ref{linear combination}) is transformed to
\begin{equation}
	\begin{aligned}
		\hat{\theta}_{true}=f^{(p)}(\hat{\boldsymbol{\Theta}}_{true})=\mathbf{W}^{(p)}\hat{\boldsymbol{\Theta}}_{true}+\mathbf{b}^{(p)},
	\end{aligned}
\end{equation}
where $\mathbf{W}^{(p)}$ and $\mathbf{b}^{(p)}$ are weight and bias vectors of this peceptron. So the proposed RegNet is the combination of the MLNN and perceptron, and the whole DOA estimation process can be expressed by
\begin{equation}
	\hat{\theta}=\mathbf{W}^{(p)}f^{(\rm{out})}(f^{(H)}(\cdots f^{(1)}(\mathbf{W}^{(\rm{in})}\hat{\boldsymbol{\Theta}}+\mathbf{b}^{(\rm{in})})))+\mathbf{b}^{(p)}.
\end{equation}

As the angle has been estimated, the next step is to estimate the range between emitter and receive array. Based on equations (\ref{c}) and (\ref{r_l2}), with the angle estimation result and the true angle solutions of all the groups, the range parameter can be obtained by 
\begin{equation}
	\hat{r}_{l_1,l_2}=\frac{\Delta d_{l_1,l_2}\cos\hat{\theta}_{l_1}\cos\hat{\theta}_{l_2}}{\cos\hat{\theta}(\sin\hat{\theta}_{l_1}\cos\hat{\theta}_{l_2}-\cos\hat{\theta}_{l_1}\sin\hat{\theta}_{l_2})}.\label{hat_r_l1_l2}
\end{equation}
Since $L$ groups can conduct $N={L\choose2}$ combinations, then the final result is the average of these $N$ estimations
\begin{equation}
	\hat{r}=\frac{1}{N}\sum_{l_1=1}^{L-1}\sum_{l_2=l_1+1}^{L}\hat{r}_{l_1,l_2}.\label{hat_r_regnet}
\end{equation}

\vspace{-1em}
\subsection{Training Strategy}
The input of the MLNN is $\hat{\boldsymbol{\Theta}}$, and the expected output is $\hat{\boldsymbol{\Theta}}_{true}$, so the training data and label pair is given as $(\hat{\boldsymbol{\Theta}},\boldsymbol{\Theta}_{true})$. And as the range of DOA is $[-\pi/2,\pi/2]$, different angles can generate different training data and label pairs, then the complete training set can be expressed by $\{(\hat{\boldsymbol{\Theta}}^{\theta},\boldsymbol{\Theta}^{\theta}_{true})|\theta\in[-\pi/2,\pi/2]\}$. In the training stage, we choose mean square error (MSE) as the loss function, which is defined as
\begin{equation}
	{\rm{MSE}}=\frac{1}{LN_d}\sum_{t=1}^{N_d}\sum_{l=1}^{L}\left(\hat{\theta}_{t,l,true}-\theta_{l,true}\right)^2,\label{MSE_MLNN}
\end{equation}
where $N_d$ denotes the number of training data and the $\theta_{l,true}$ is the ideal output of the $l$th output neuron. Then by minimizing (\ref{MSE_MLNN}) the optimal parameters of the proposed MLNN can be obtained. Similarly, the training set of the perceptron is $\{(\hat{\boldsymbol{\Theta}}^{\theta}_{true},\theta)|\theta\in[-\pi/2,\pi/2]\}$, and the loss function is also MSE:
\begin{equation}
	{\rm{MSE}}_p=\frac{1}{N_d}\sum_{t=1}^{N_d}\left(\hat{\theta}_{t,true}-\theta\right)^2.\label{MSE_p}
\end{equation}
Obviously, if we only want to obtain the angle information, theses two parts can be combined as an ensemble for reducing training complexity.

\vspace{-1em}
\section{Discussion and Analysis}\label{performance}
\subsection{Muilti-emitter Scenario Discussion}
Although this paper focuses on a single-emitter scenario for presenting the principles of the investigated problem more clearly, the proposed structure and algorithms in this paper are also applicable to multi-emitter scenario, then we will make some analysis on multi-emitter situation in this subsection. Assuming there are $Q$ emitters, then the DOA estimation problem within each group that solved by Root-MUSIC algorithm is equivalent to evaluating $Q$ roots that be closest to unit circle, and each group can generate a initial set based on each root. So the obtained initial set in (\ref{hat_Theta_l}) becomes $\hat{{\Theta}}_{l,1},\hat{{\Theta}}_{l,2},\cdots,\hat{{\Theta}}_{l,Q}$. After that, perform angle calibration on the obtained initial sets, we can get $Q$ candidate position sets that corresponding to $Q$ emitters, and each candidate position set has the form of (\ref{omega}). Therefore, the proposed ML-based algorithms can be utilized on the $Q$ candidate position sets respectively, and the $Q$-emitter localization problem is transformed to $Q$ independent single-emitter problems. Anyway, under the proposed NF emitter localization framework in this paper, the multi-emitter problem can be viewed as multiple single-emitter problems, so we only need to focus on the single-emitter case.
\vspace{-2.5em}
\subsection{Complexity Analysis}
From Fig. \ref{subarray method} we can know the proposed methods share the common computation complexity from the Root-MUSIC algorithm in the FF DOA estimation stage, and have different computation complexities for the NF emitter localization. The computation complexity of Root-MUSIC comes from covariance matrix computation, SVD and signal subspace extraction, it can be approximately given as $\mathcal{O}(L(TG^2+G^3))$, where $L=K/G$. As a key step of clustering-based methods, the complexity of angle calibration is $\mathcal{O}(2NMs)$. Then the computation complexity of the proposed MSDC in the NF emitter localization stage only comes from computing the sample distance, so it is $\mathcal{O}(M_sN(N-1)/2)$. For the proposed RSD-ASD-DBSCAN, its complexity depends on the number of samples and how many iterations are required to find the only cluster that has $N$ samples, so it is given as $\mathcal{O}(\sum_{j=1}^{J}N_{s,j}^2)$, where $N_{s,j}>N$ denotes the number of samples to be divided at $j$th iteration. Finally, the complexity of the proposed RegNet is related to the size of the MLNN, i.e., the depth and the number of neurons contained in each hidden layer. Then it is approximately given as $\mathcal{O}(N_eL^2M_s\prod_{h=1}^{H}N_h)$, where $N_e$ denotes the number of epochs and $N_h$ is the neuron number of hidden layers.

\vspace{-1.5em}
\subsection{CRLB Analysis}
The CRLB provides a lower bound on the variance of any unbiased estimator. 
So the CRLB of the grouped PC-HAD architecture is derived in this section as a benchmark to evaluate the NF localization performance of the proposed methods. Since the signal model (\ref{FF_model}) is only related to angle parameter, we have to first transform it to NF model. Based on the geometric relationship depicted in Fig.\ref{angle} we get
\begin{equation}
	\begin{cases}
	r_l\cos\theta_l=r\cos\theta\\
	r_l\sin\theta_l=r\sin\theta-\Delta d_l
	\end{cases},
\end{equation}
where $l=1,2,\cdots,L$ and $\Delta d_l=(l-1)GM_sd$. So we can express $\sin\theta_l$ by
\begin{equation}
	\varphi_l=\sin\theta_l=\frac{r\sin\theta-\Delta d_l}{\sqrt{r^2-2\Delta d_lr\sin\theta+\Delta d_l^2}},
\end{equation}
and the steering vector (\ref{FF array steering vector}) also becomes
$\mathbf{a}(\varphi_l)=\mathbf{a}_g(\varphi_l)\otimes\mathbf{a}_s(\varphi_l)$,
where $\mathbf{a}_s(\varphi_l)=[1, e^{j\frac{2\pi d}{\lambda}\varphi_l},\cdots,e^{j\frac{2\pi d}{\lambda}(M_s-1)\varphi_l}]^T$, and $\mathbf{a}_g(\theta_l)=[1, e^{j\frac{2\pi d}{\lambda}M_s\varphi_l},\cdots,e^{j\frac{2\pi d}{\lambda}(G-1)M_s\varphi_l}]^T$. So the covariance matrix of group $l$ is given as 
\begin{equation}
	\mathbf{R}_l=\sigma_s^2\mathbf{a}_W(\varphi_l)\mathbf{a}_W^H(\varphi_l)+\sigma_v^2\mathbf{I}_{G},\label{R_l}
\end{equation}
where $\mathbf{a}_W(\varphi_l)=\mathbf{W}_l^H\mathbf{a}(\varphi_l)$. Then by combining all the groups we can get
\begin{equation}
	\mathbf{R}={\rm{diag}}\{\mathbf{R}_1,\mathbf{R}_2,\cdots,\mathbf{R}_L\},\label{group_R}
\end{equation}
and the CRLB can be derived based on it.

\textit{Lemma 1:} The CRLBs for the DOA and range estimations of grouped HAD architecture can be expressed by
\begin{subequations}
	\begin{align}
		{\rm{CRLB}}_{\theta}=\frac{\sum_{l=1}^{L}F_{rr}^{(l)}}{T\left[\sum_{l=1}^{L}F_{\theta\theta}^{(l)}\sum_{l=1}^{L}F_{rr}^{(l)}-(\sum_{l=1}^{L}F_{\theta r}^{(l)})^2\right]}\\
		{\rm{CRLB}}_{r}=\frac{\sum_{l=1}^{L}F_{\theta\theta}^{(l)}}{T\left[\sum_{l=1}^{L}F_{\theta\theta}^{(l)}\sum_{l=1}^{L}F_{rr}^{(l)}-(\sum_{l=1}^{L}F_{\theta r}^{(l)})^2\right]}
	\end{align}
\end{subequations}
where $F_{rr}^{(l)}$, $F_{\theta\theta}^{(l)}$ and $F_{\theta r}^{(l)}$ are defined in (\ref{F_rr}), (\ref{F_tt}) and (\ref{F_tr}).

\textit{Proof:} See Appendix A. $\hfill\blacksquare$

By observing the expression of CRLB, as the basic HAD structure parameters $M_s$ and $K$ are fixed, the structure of grouped PC-HAD array is only related to $G$, i.e., the number of subarrays in each group. Therefore, by analyzing its impact on CRLB, we get a remark as follow:

\textit{Remark 1:} The value of CRLB decreases with the increasing of $G$.

\textit{Proof:} See Appendix B. $\hfill\blacksquare$

\begin{table}[tb]
	\centering
	\caption{Computation Complexity of Proposed Methods}
		\begin{tabular}{c|c}
			\Xhline{1.0pt}
			Methods & Complexity\\
			\Xhline{0.5pt}
			MSDC & $\mathcal{O}(TKG+KG^2+2NMs+M_sN(N-1)/2)$ \\
			\Xhline{0.5pt}
			RSD-ASD-DBSCAN & $\mathcal{O}(TKG+KG^2+2NMs+\sum_{j=1}^{J}N_{s,j}^2)$ \\
			\Xhline{0.5pt}
			RegNet & $\mathcal{O}(TKG+KG^2+N_eL^2M_s\prod_{h=1}^{H}N_h)$ \\
			\Xhline{1.0pt}
		\end{tabular}\label{table_complexity}
	\vspace{-1em}
\end{table}

\section{Simulation Results}
In this section, we present some simulation results to evaluate the performances of the three proposed localization methods: MSDC, RSD-ASD-DBSCAN and RegNet. The CRLB of the proposed grouped HAD structure is also considered as benchmark. Then the receive array employed in simulations is a ULA with $M=240$, $M_s=3$ and $L=5$, carrier frequency $f_c=30\rm{GHz}$, the emitter position is set as $(\theta,r)=(60^{\circ},20m)$. Under this condition, the MLNN part of the proposed RegNet is a fully-connected network has $LM_s=15$ input neurons, two hidden layers with 12 and 8 neurons respectively, and also has $L=5$ output neurons. The RegNet is trained for 100 epochs and optimized by Adam. The localization accuracy of the proposed methods are evaluated by the root mean-squared error (RMSE) of DOA and range estimations, and all results are averaged over 10000 Monte-Carlo simulations.

\begin{figure}[t]
	\centering
	\subfigure[]{\includegraphics[width=0.37\textwidth]{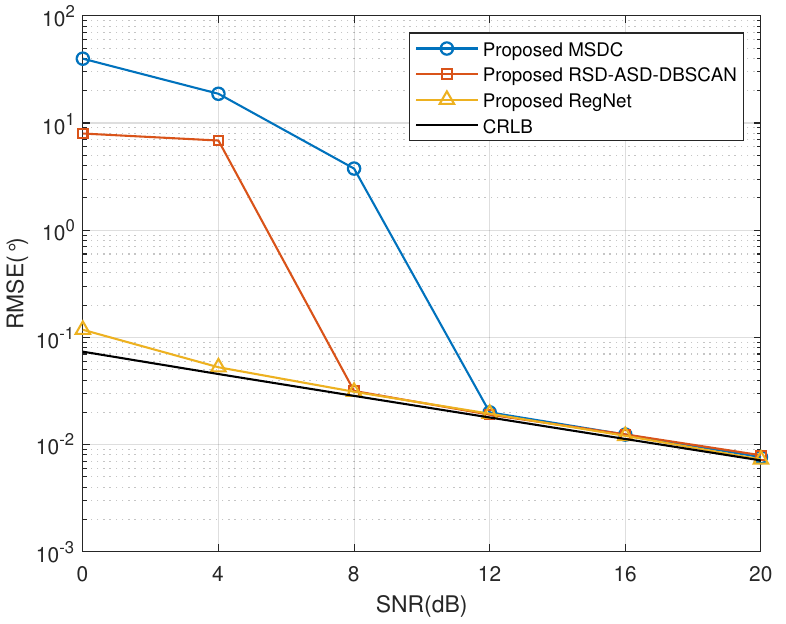}}\\
	\subfigure[]{\includegraphics[width=0.37\textwidth]{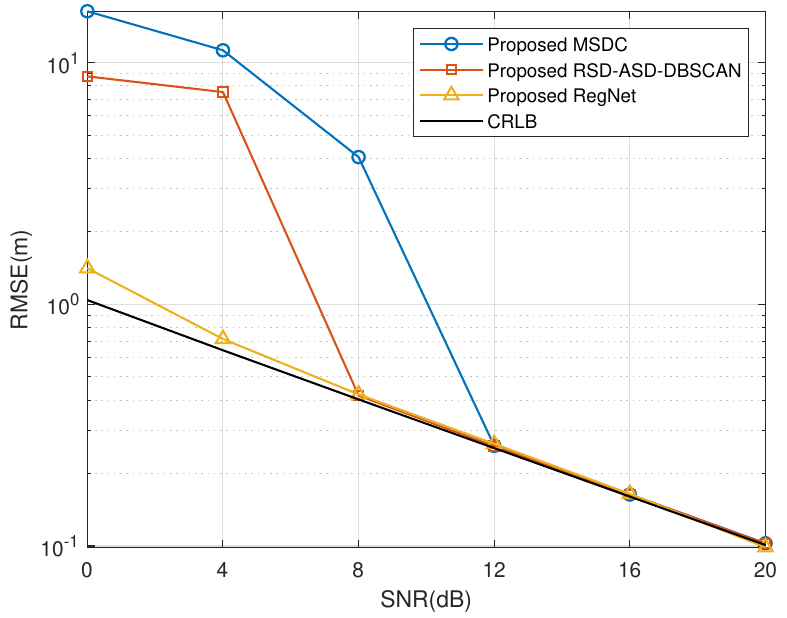}}
	\caption{RMSE versus SNR, (a) DOA estimation, (b) range estimation.}\label{rmse_snr}
	\vspace{-2em}
\end{figure}

\begin{figure}[h]
	\centering
	\subfigure[]{\includegraphics[width=0.37\textwidth]{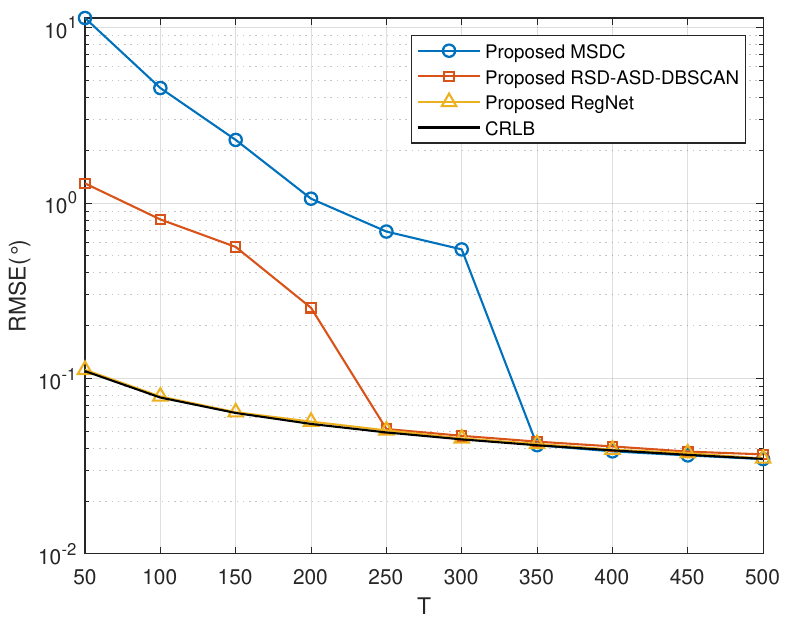}}\\
	\subfigure[]{\includegraphics[width=0.37\textwidth]{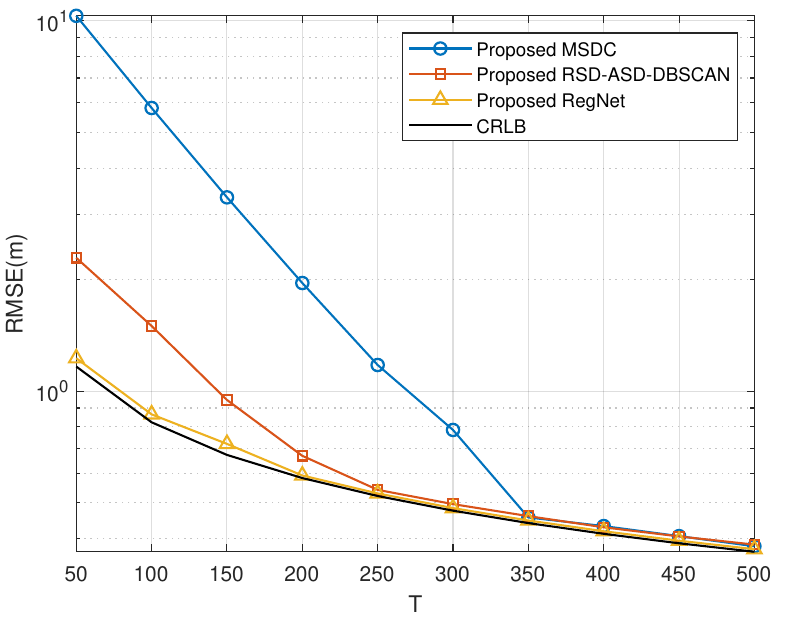}}
	\caption{RMSE versus the number of snapshots $T$, (a) DOA estimation, (b) range estimation.}\label{rmse_T}
	\vspace{-1em}
\end{figure}

Fig.\ref{rmse_snr} plots the RMSE of DOA and range estimations versus SNR for the proposed three methods and CRLB with $T=100$. Since the emitter localization results of the proposed methods are all based on the DOA estimation results obtained by Root-MUSIC algorithm, which is an unbiased estimator, then the estimation performance of the proposed methods can achieve the CRLB. By combining (a) and (b), it is clear that the proposed RegNet outperforms other methods, its DOA and range estimation accuracy can both achieve the CRLB at $\rm{SNR}=8dB$. Then based on the performance of DOA estimation and distance estimation respectively, compared to the two clustering-based methods, the proposed RegNet has a significant localization performance advantage in the low SNR region, i.e., $\rm{SNR}<8dB$. RSD-ASD-DBSCAN is a method with performance second only to the RegNet, its estimation accuracy can achieve CRLB at $\rm{SNR}=8dB$, and its localization performance evidently outperforms MSDC in the medium-to-low SNR region. Finally, MSDC has a good localization performance with high SNR, it can achieve CRLB for DOA and range estimations at $\rm{SNR}=12dB$. 

Fig.\ref{rmse_T} illustrates the localization accuracy of the proposed methods versus the number of snapshots $T$ with $\rm{SNR}=6dB$. Similar to the trend of RMSE versus SNR, the proposed RegNet has great advantages over RSD-ASD-DBSCAN and MSDC respectively at region of $T<250$ and $T<350$, it also achieves CRLB at $T=50$. RSD-ASD-DBSCAN performs significantly better than MSDC as $T<350$ and attains CRLB at $T= 250$. MSDC also achieves CRLB at $T=350$. Therefore, by combining Fig.\ref{rmse_snr} and Fig.\ref{rmse_T}, we can find the proposed RegNet has robust localization performance in the low SNR and number of snapshots situations, while clustering-based methods need higher conditions to achieve the same performance as RegNet.

\begin{figure}[t]
	\centering
	\subfigure[]{\includegraphics[width=0.365\textwidth]{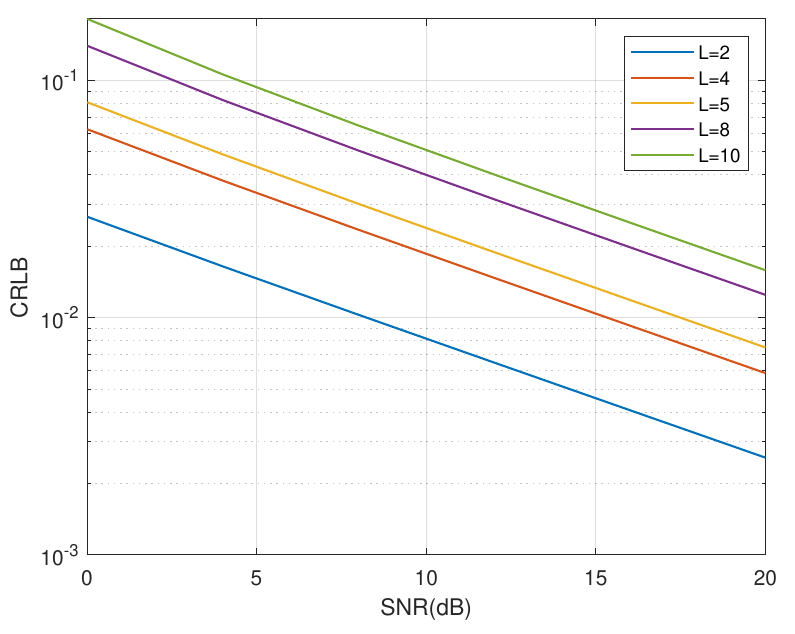}}\\
	\subfigure[]{\includegraphics[width=0.37\textwidth]{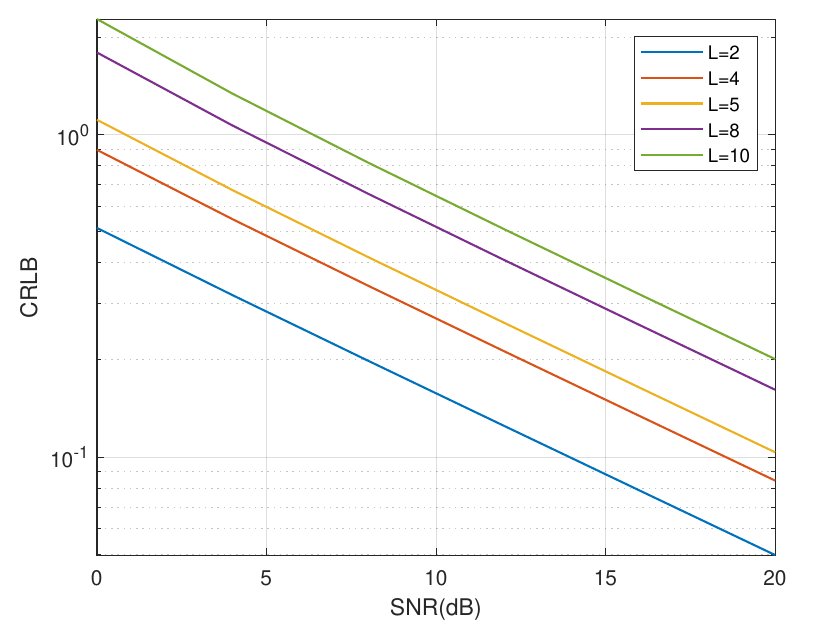}}
	\caption{CRLB versus SNR, (a) DOA estimation, (b) range estimation.}\label{CRLB}
	\vspace{-1.4em}
\end{figure}

As with the fixed array size $M$ and the number of subarrays $K$, the structure of grouped HAD array depends on the number of groups $L$. So in order to explore the impact of array structures on the localization performance, we plot the CRLB versus SNR for grouped PC-HAD arrays with different $L$. In figure (a), the comparison is made between the grouped PC-HAD arrays with different $L$ in DOA estimation, we can see the grouped arrays with less number of groups have lower CRLB which mean they have potential to achieve more accurate localization results. Fig.\ref{CRLB} (b) compares the CRLB of different $L$ in range estimation, and it shows the similar trend as DOA estimation. Since $K=LG$, the relationship between CRLB and $L$ depicted in this simulation also accords with remark 1 in the section \ref{performance}.

\begin{figure}[t]
	\centering
	\subfigure[]{\includegraphics[width=0.37\textwidth]{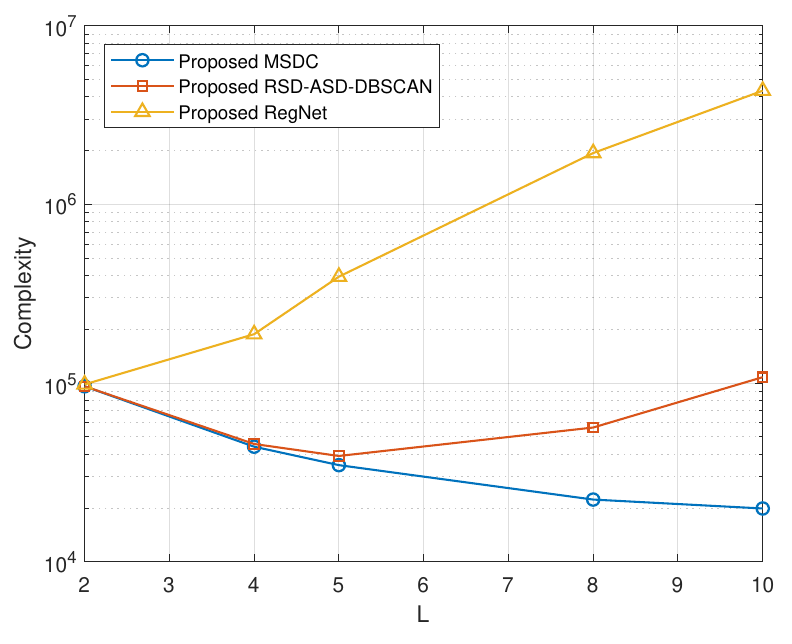}}\\
	\subfigure[]{\includegraphics[width=0.37\textwidth]{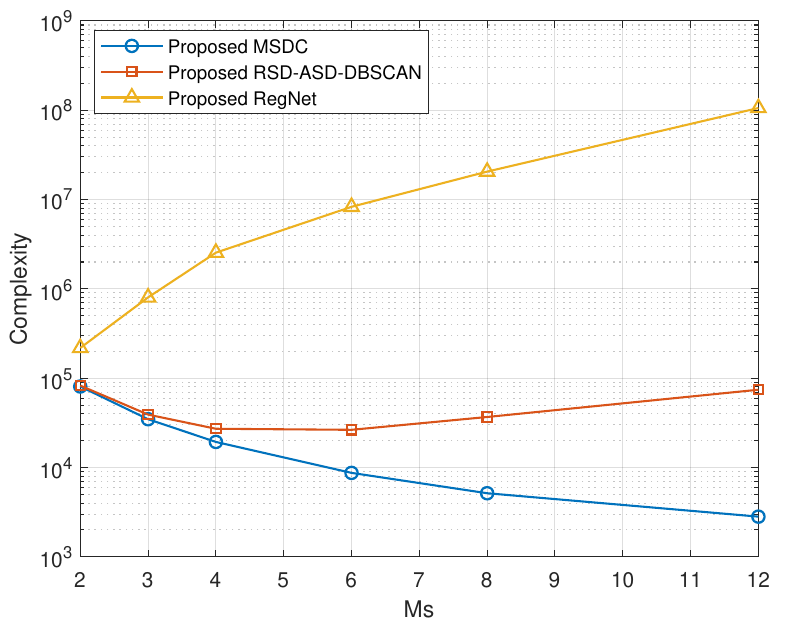}}
	\caption{Computation complexity comparison, (a) complexity versus $L$ with $M_s=3$, (b) complexity versus $M_s$ with $L=5$.}\label{complexity}
	\vspace{-1em}
\end{figure}

Fig.\ref{complexity} depicts the computation complexity of the proposed methods versus $L$ and $M_s$. The specific complexity expressions are given in the Table \ref{table_complexity}. Since the input size of the proposed RegNet is $LM_s$, and network size including depth and hidden layer size is strongly associated with the input size, the increasing of $L$ or $M_s$ will lead to the enlargement of RegNet size. So the complexity of RegNet increases apparently with the growth of $L$ and $M_s$, and is much higher than RSD-ASD-DBSCAN and MSDC for the unsupervised methods can save the computation spending of training stage. The sample number of the proposed RSD-ASD-DBSCAN also raises with the increasing of $L$ and $M_s$, its complexity exceeds that of MSDC at $L=4$ and $M_s=3$, and the difference becomes larger with the growth of $L$ and $M_s$.

\vspace{-1em}
\section{Conclusion}
	In this paper, we first proposed a grouped PC-  HAD structure for NF DOA estimation. By dividing the large-scale receive into several small-scale groups, the NF problem of estimating DOA within each group is viewed as a FF one. Then a candidate position set was generated by calibrating the estimation results of all the groups to a common reference point. Based on the distribution characters of the samples in the candidate position set, two low-complexity clustering-based methods, MSDC and RSD-ASD-DBSCAN, were proposed to solve phase ambiguity problem and locate the NF emitter. In order to further improve the localization accuracy, we proposed a high-resolution RegNet as well, which contains an MLNN for false solution elimination and a perceptron for angle fusion. Eventually, the CRLB of NF emitter localization for proposed grouped PC-HAD structure was derived as a performance benchmark for the proposed methods. The simulation results showed that the clustering-based methods can achieve the CRLB at high SNR region with extremely low complexity, while the proposed RegNet attains CRLB at much lower SNR level and has significantly performance improvement in the medium-to-low SNR regions.  
\vspace{-1em}
\appendices
\section{Derivation of CRLB} 
The Fisher information matrix (FIM) of the NF localization problem is
\begin{equation}
	\mathbf{F}=\begin{bmatrix}
		F_{\theta\theta} & F_{\theta r}\\
		F_{r\theta} & F_{rr}
	\end{bmatrix},\label{FIM}
\end{equation}
then based on (\ref{group_R}), the expression of elements in $\mathbf{F}$ is given as
\begin{equation}
		F_{\beta_i\beta_j}={\rm{tr}}\left\{\mathbf{R}^{-1}\frac{\partial \mathbf{R}}{\partial \beta_i}\mathbf{R}^{-1}\frac{\partial \mathbf{R}}{\partial \beta_j}\right\}=\sum_{l=1}^{L}F_{\beta_i\beta_j}^{(l)},
\label{FIM_element}
\end{equation}
where $\boldsymbol{\beta}=[\theta,r]^T$, $1\leq i,j\leq 2$ and $F_{\beta_i\beta_j}^{(l)}={\rm{tr}}\left\{\mathbf{R}_l^{-1}\frac{\partial \mathbf{R}_l}{\partial \beta_i}\mathbf{R}_l^{-1}\frac{\partial \mathbf{R}_l}{\partial \beta_j}\right\}$.
Then referring to (\ref{R_l}), we can first attain two basic terms of $F_{\beta_i\beta_j}^{(l)}$
\begin{equation}
	\begin{aligned}
		\frac{\partial \mathbf{R}_l}{\partial \theta}=\mathbf{W}_l^H\left(\dot{\mathbf{a}}_{l,\theta}\mathbf{a}_l^H+\mathbf{a}_l\dot{\mathbf{a}}^H_{l,\theta}\right)\mathbf{W}_l,\\
		\frac{\partial \mathbf{R}_l}{\partial r}=\mathbf{W}_l^H\left(\dot{\mathbf{a}}_{l,r}\mathbf{a}_l^H+\mathbf{a}_l\dot{\mathbf{a}}^H_{l,r}\right)\mathbf{W}_l,
	\end{aligned}\label{derivation}
\end{equation}
where $\mathbf{a}_l=\mathbf{a}(\varphi_l)$ and
\begin{equation}
	\begin{aligned}
		\dot{\mathbf{a}}_{l,\theta}=\frac{\partial \mathbf{a}_l}{\partial \theta}=j\frac{2\pi d}{\lambda}\dot{\varphi}_{l,\theta}\mathbf{D}\mathbf{a}_l,\quad \dot{\mathbf{a}}_{l,r}=\frac{\partial \mathbf{a}_l}{\partial r}=j\frac{2\pi d}{\lambda}\dot{\varphi}_{l,r}\mathbf{D}\mathbf{a}_l,
	\end{aligned}
\end{equation}
where $\mathbf{D}={\rm{diag}}\{0, 1,\cdots,GM_s-1\}$, and
\begin{equation}
	\begin{aligned}
		\dot{\varphi}_{l,\theta}=\frac{\partial \varphi_l}{\partial \theta}=\frac{r^3\cos\theta-\Delta d_l r^2\sin\theta\cos\theta}{(\sqrt{r^2-2\Delta d_lr\sin\theta+\Delta d_l^2})^3},\\
		\dot{\varphi}_{l,r}=\frac{\partial \varphi_l}{\partial r}=\frac{\Delta d_lr\cos^2\theta}{(\sqrt{r^2-2\Delta d_lr\sin\theta+\Delta d_l^2})^3}.\label{varphi}
	\end{aligned}
\end{equation}
So (\ref{derivation}) can be further expressed by 
\begin{equation}
	\begin{aligned}
		\frac{\partial \mathbf{R}_l}{\partial \theta}=j\frac{2\pi d}{\lambda}\dot{\varphi}_{l,\theta}\mathbf{W}_l^H\left(\mathbf{D}\mathbf{a}_l\mathbf{a}_l^H-\mathbf{a}_l\mathbf{a}_l^H\mathbf{D}\right)\mathbf{W}_l,\\
		\frac{\partial \mathbf{R}_l}{\partial r}=j\frac{2\pi d}{\lambda}\dot{\varphi}_{l,r}\mathbf{W}_l^H\left(\mathbf{D}\mathbf{a}_l\mathbf{a}_l^H-\mathbf{a}_l\mathbf{a}_l^H\mathbf{D}\right)\mathbf{W}_l,
	\end{aligned}
\end{equation}
and by substituting it into $F_{\beta_i\beta_j}^{(l)}$ we can get
\begin{subequations}
	\begin{align}
		&F_{\theta\theta}^{(l)}=\dot{\varphi}_{l,\theta}^2\mu^2{\rm{tr}}\left\{\left[\mathbf{R}_l^{-1}\mathbf{W}_l^H\left(\mathbf{D}\mathbf{a}_l\mathbf{a}_l^H-\mathbf{a}_l\mathbf{a}_l^H\mathbf{D}\right)\mathbf{W}_l\right]^2\right\}\\
		&F_{rr}^{(l)}=\dot{\varphi}_{l,r}^2\mu^2{\rm{tr}}\left\{\left[\mathbf{R}_l^{-1}\mathbf{W}_l^H\left(\mathbf{D}\mathbf{a}_l\mathbf{a}_l^H-\mathbf{a}_l\mathbf{a}_l^H\mathbf{D}\right)\mathbf{W}_l\right]^2\right\}\\
		&F_{\theta r}^{(l)}=\dot{\varphi}_{l,\theta}\dot{\varphi}_{l,r}\mu^2{\rm{tr}}\left\{\left[\mathbf{R}_l^{-1}\mathbf{W}_l^H\left(\mathbf{D}\mathbf{a}_l\mathbf{a}_l^H-\mathbf{a}_l\mathbf{a}_l^H\mathbf{D}\right)\mathbf{W}_l\right]^2\right\}\\
		&F_{r\theta}^{(l)}=\dot{\varphi}_{l,r}\dot{\varphi}_{l,\theta}\mu^2{\rm{tr}}\left\{\left[\mathbf{R}_l^{-1}\mathbf{W}_l^H\left(\mathbf{D}\mathbf{a}_l\mathbf{a}_l^H-\mathbf{a}_l\mathbf{a}_l^H\mathbf{D}\right)\mathbf{W}_l\right]^2\right\}
	\end{align}\label{F_l}
\end{subequations}
where $\mu=j2\pi d/\lambda$. 
Then referring to the CRLB of the PC-HAD structure derived in \cite{shu2018low}, and substituting (\ref{varphi}) into the above equation, $F_{\beta_i\beta_j}^{(l)}$ are further expressed by
\begin{subequations}
	\begin{align}
		&F_{\theta \theta}^{(l)}=\frac{\mu^2r^4\cos^2\theta(r-\Delta d_l\sin\theta)^2}{(r^2-2\Delta d_lr\sin\theta+\Delta d_l^2)^3}\Xi_l,\label{F_tt}\\
		&F_{rr}^{(l)}=\frac{\mu^2\Delta d_l^2r^2\cos^4\theta}{(r^2-2\Delta d_lr\sin\theta+\Delta d_l^2)^3}\Xi_l,\label{F_rr}\\
		&F_{\theta r}^{(l)}=F_{r\theta}^{(l)}=\frac{\mu^2\Delta d_lr^3\cos^3\theta(r-\Delta d_l\sin\theta)}{(r^2-2\Delta d_lr\sin\theta+\Delta d_l^2)^3}\Xi_l,\label{F_tr}
	\end{align}\label{F_l_new}
\end{subequations}
where $\Xi_l$ is given by (\ref{F_l_ij}), $\gamma=\sigma_s^2/\sigma_v^2$, and
\begin{figure*}[t]
	\begin{equation}
			\Xi_l
			=-\frac{2\gamma^2}{d^2M_s(M_s+\gamma G\lVert\Gamma_l\rVert^2)^2}\left[\frac{\lVert\Gamma_l\rVert^4M_s^2G^2(G^2-1)(M_s+\gamma G\lVert\Gamma_l\rVert^2)}{12}+M_sG\lVert\Gamma_l\rVert^2\lVert\eta_l\rVert^2+M_sG^2\Re(\Gamma_l^2\eta_l)\right]\label{F_l_ij}
	\end{equation}
\hrulefill
\end{figure*}

\begin{subequations}
	\begin{align}
		&\Gamma_l=\sum_{m_s=1}^{M_s}e^{j\frac{2\pi d}{\lambda}(m_s-1)\varphi_l}=\frac{1-e^{j\frac{2\pi d}{\lambda}M_s\varphi_l}}{1-e^{j\frac{2\pi d}{\lambda}\varphi_l}},\\
		&\eta_l=\sum_{m_s=1}^{M_s}(m_s-1)e^{-j\frac{2\pi d}{\lambda}(m_s-1)\varphi_l}.
	\end{align}
\end{subequations}

As CRLB is the inversion of FIM, given the number of snapshots $T$, then the CRLB of the proposed grouped hybrid architecture for DOA and range estimations are given by
\begin{equation}
	\begin{aligned}
		{\rm{CRLB}}_{\theta}
		=\frac{\sum_{l=1}^{L}F_{rr}^{(l)}}{T\left[\sum_{l=1}^{L}F_{\theta\theta}^{(l)}\sum_{l=1}^{L}F_{rr}^{(l)}-(\sum_{l=1}^{L}F_{\theta r}^{(l)})^2\right]},
	\end{aligned}\label{crlb_theta}  
\end{equation}
\begin{equation}
	\begin{aligned}
		{\rm{CRLB}}_{r}
		=\frac{\sum_{l=1}^{L}F_{\theta\theta}^{(l)}}{T\left[\sum_{l=1}^{L}F_{\theta\theta}^{(l)}\sum_{l=1}^{L}F_{rr}^{(l)}-(\sum_{l=1}^{L}F_{\theta r}^{(l)})^2\right]}.
	\end{aligned}\label{crlb_r}  
\end{equation}

\vspace{-1.5em}
\section{Proof of Remark 1}
By observing the expressions of (\ref{F_l}) and (\ref{F_l_new}), we can get $F_{\theta r}^{(l)}=\sqrt{F_{\theta\theta}^{(l)}F_{rr}^{(l)}}$. Therefore, let $F_{\theta\theta}^{(l)}=F_{\theta,l}^2$ and $F_{rr}^{(l)}=F_{r,l}^2$, thus $F_{\theta r}^{(l)}=F_{\theta,l}F_{r,l}$. Then based on Cauchy-Schwarz inequality, we can get the inequalities as (\ref{inequality}). The numerator and denominator of ${\rm{CRLB}}_{\theta}$ are both divided by $F_{rr}$, the new expression of ${\rm{CRLB}}_{\theta}$ is given as
\begin{equation}
	{\rm{CRLB}}_{\theta}=\frac{1}{T(F_{\theta\theta}-F_{\theta r}^2/F_{rr})},
\end{equation}
so the problem is equivalent to evaluating the monotonicity of $F'=F_{\theta\theta}-F_{\theta r}^2/F_{rr}$ with respect to $G$. 

Assuming the value of $F'$ increases with the growth of $G$, given $G_2>G_1$ and we can get
\begin{equation}
	\begin{aligned}
		&F'(G_2)-F'(G_1)\geq 0\Rightarrow (\ref{F2-F1})\\		
		\Rightarrow &F_{\theta\theta}(G_2)F_{rr}(G_2)F_{rr}(G_1)\geq F_{\theta r}^2(G_2)F_{rr}(G_1),
	\end{aligned}
\end{equation}
based on (\ref{inequality}) we know the above inequality holds true, so the assumption is valid. Thus the value of ${\rm{CRLB}}_{\theta}$ will decrease with respect to $G$, and the trend of ${\rm{CRLB}}_{r}$ is identical.

\begin{figure*}[t]
	\begin{equation}
		\left(F_{\theta,1}^2+\cdots+F_{\theta,L}^2\right)\left(F_{r,1}^2+\cdots+F_{r,L}^2\right)\geq \left(F_{\theta,1}F_{r,1}+\cdots+F_{\theta,L}F_{r,L}\right)^2\Longleftrightarrow F_{\theta\theta}F_{rr}-F_{\theta r}^2\geq 0\label{inequality}
	\end{equation}
	\hrulefill
\end{figure*}
\begin{figure*}[t]
	\begin{equation}
		\begin{aligned}
			F_{\theta r}^2(G_1)F_{rr}(G_2)-F_{\theta r}^2(G_2)F_{rr}(G_1)&\geq F_{\theta\theta}(G_1)F_{rr}(G_1)F_{rr}(G_2)-F_{\theta\theta}(G_2)F_{rr}(G_1)F_{rr}(G_2)\\
			&\geq F_{\theta r}^2(G_1)F_{rr}(G_2)-F_{\theta\theta}(G_2)F_{rr}(G_1)F_{rr}(G_2)
		\end{aligned}\label{F2-F1}
	\end{equation}
	\hrulefill
\end{figure*}

	\bibliographystyle{IEEEtran}
	\bibliography{hybrid}	
\end{document}